\documentclass[reqno]{siamart190516}
\usepackage{etex}
\usepackage{enumerate}
\usepackage{algorithm,algorithmic}
\usepackage{epic,eepic,graphicx}
\usepackage{multicol}
\usepackage{subcaption}
\usepackage{epstopdf}
\usepackage{amsmath,amssymb}
\usepackage{pgfplots}
\usepackage{bbm}
\usepackage{booktabs}
\usepackage{tabularx}
\usepackage{color}
\usepackage{colortbl}
\usepackage{pgfplotstable}
\usepackage{tikz}
\usepackage{rotating}
\usepackage{multirow,bigdelim}
\usepackage{verbatim}
\usepackage{lscape}
\usetikzlibrary{pgfplots.groupplots}
\usetikzlibrary{matrix,positioning}
\usetikzlibrary{external}
\tikzsetexternalprefix{Tikz/}
\usetikzlibrary{shapes}
\usetikzlibrary{arrows}
\usetikzlibrary{calc}  
    \usetikzlibrary{
	pgfplots.colorbrewer,
}
\pgfplotsset{compat=newest}
\pgfplotsset{
	cycle list/.define={my marks}{
		every mark/.append style={solid,fill=\pgfkeysvalueof{/pgfplots/mark list fill}},mark=*\\
		every mark/.append style={solid,fill=\pgfkeysvalueof{/pgfplots/mark list fill}},mark=square*\\
		every mark/.append style={solid,fill=\pgfkeysvalueof{/pgfplots/mark list fill}},mark=triangle*\\
		every mark/.append style={solid,fill=\pgfkeysvalueof{/pgfplots/mark list fill}},mark=diamond*\\
	},
}

\pgfplotscreateplotcyclelist{colors}{%
	red,blue,teal,orange,cyan,green!70!black,black,violet,magenta,gray,yellow,brown}
\pgfplotscreateplotcyclelist{clusters}{%
	blue,thick,every mark/.append style={solid,fill opacity=0.7},mark=*\\%
	red,thick,every mark/.append style={solid,fill opacity=0.7},dash dot,mark=square*\\%
	black!40!green,thick,every mark/.append style={solid,fill opacity=0.7},mark=triangle*\\%
	black!20!cyan,thick,every mark/.append style={solid,fill opacity=0.7},dashed,mark=diamond*\\%
	black!20!magenta,thick,every mark/.append style={solid,fill opacity=0.7 },mark=pentagon*\\%
	gray,densely dotted,every mark/.append style={solid,fill opacity=0.7},mark=otimes*\\%
	cyan,dashed,every mark/.append style={solid,fill opacity=0.7},mark=oplus*\\%
	red,every mark/.append style={fill=blue!80!black},mark=*\\%
	black!40!red,densely dotted,every mark/.append style={fill=blue!80!black},mark=*\\%
	black!20!red,dashed,every mark/.append style={fill=blue!80!black},mark=*\\%
	green,every mark/.append style={fill=blue!80!black},mark=*\\%
	black!40!green,densely dotted,every mark/.append style={fill=blue!80!black},mark=*\\%
	black!20!green,dashed,every mark/.append style={fill=blue!80!black},mark=*\\%
	cyan,every mark/.append style={fill=blue!80!black},mark=*\\%
	black!40!cyan,densely dotted,every mark/.append style={fill=blue!80!black},mark=*\\%
	black!20!cyan,dashed,every mark/.append style={fill=blue!80!black},mark=*\\%
}
\pgfplotscreateplotcyclelist{smoothers}{%
	blue,thick,,every mark/.append style={fill=blue!80!black},mark=*\\%
	red,thick,dash dot,every mark/.append style={fill=blue!80!black},mark=*\\%
	black!40!green,thick,densely dotted,every mark/.append style={fill=blue!80!black},mark=*\\%
	black!20!cyan,thick,dashed,every mark/.append style={fill=blue!80!black},mark=*\\%
	blue,every mark/.append style={fill=blue!80!black},mark=*\\%
	red,every mark/.append style={fill=blue!80!black},mark=*\\%
	green,every mark/.append style={fill=blue!80!black},mark=*\\%
	black!40!blue,densely dotted,every mark/.append style={fill=blue!80!black},mark=*\\%
	black!40!red,densely dotted,every mark/.append style={fill=blue!80!black},mark=*\\%
	black!40!green,densely dotted,every mark/.append style={fill=blue!80!black},mark=*\\%
	black!20!blue,dashed,every mark/.append style={fill=blue!80!black},mark=*\\%
	black!20!red,dashed,every mark/.append style={fill=blue!80!black},mark=*\\%
	black!20!green,dashed,every mark/.append style={fill=blue!80!black},mark=*\\%
}

\newlength\myindent
\usepackage{aeguill}

\tikzset{class or interface/.style={%
		draw,%
		shape=rectangle split,%
		rectangle split parts=3,%
		rectangle split part align={center,left,left},%
		rectangle split part fill={#1!30!white,#1!20!white,#1!10!white},
		every two node part/.style={align=center, font=\ttfamily},
		every three node part/.style={align=left, font=\ttfamily}, 
		node distance=2
	},
	interface/.style={
		class or interface=blue!80!black,
		every one node part/.style={align=center, font=\ttfamily\itshape}
	},                       
	concrete class/.style={
		class or interface=green!70!black,
		every one node part/.style={align=center, font=\ttfamily\bfseries}
	}
}

\pgfplotsset{
	log x ticks with fixed point/.style={
		xticklabel={
			\pgfkeys{/pgf/fpu=true}
			\pgfmathparse{exp(\tick)}%
			\pgfmathprintnumber[fixed relative, precision=3]{\pgfmathresult}
			\pgfkeys{/pgf/fpu=false}
		}
	},
	log y ticks with fixed point/.style={
		yticklabel={
			\pgfkeys{/pgf/fpu=true}
			\pgfmathparse{exp(\tick)}%
			\pgfmathprintnumber[fixed relative, precision=3]{\pgfmathresult}
			\pgfkeys{/pgf/fpu=false}
		}
	}
}

\usepackage{lipsum}
\usepackage{amsfonts}
\usepackage{graphicx}
\usepackage{epstopdf}
\usepackage{algorithmic}
\ifpdf
  \DeclareGraphicsExtensions{.eps,.pdf,.png,.jpg}
\else
  \DeclareGraphicsExtensions{.eps}
\fi

\newcommand{\TheTitle}{Multilevel Gibbs Sampling for Bayesian regression} 
\newcommand{\TheAuthors}{J. Tavernier, J. Simm, \'A. Arany, K. Meerbergen and Y. Moreau}

\headers{\TheTitle}{\TheAuthors}

\title{{\TheTitle}\thanks{This work was supported Research Foundation - Flanders (FWO, No. G079016).}}

\author{
  Joris Tavernier\thanks{Departement of Computer Science, KU Leuven,
    \email{Joris.Tavernier@cs.kuleuven.be}, \url{https://people.cs.kuleuven.be/\~joris.tavernier/}, \email{Karl.Meerbergen@cs.kuleuven.be}.}
  \and
  Jaak Simm\thanks{Department of Electrical Engineering, ESAT - STADIUS, KU Leuven, \email{jaak.simm@esat.kuleuven.be}, \email{adam.arany@esat.kuleuven.be}, \email{moreau@esat.kuleuven.be}.}
   \and
   \'Ad\'am Arany\footnotemark[3]
 \and
 Karl Meerbergen\footnotemark[2]
 \and
 Yves Moreau\footnotemark[3]
}

\usepackage{amsopn}

\ifpdf
\hypersetup{
	pdftitle={\TheTitle},
	pdfauthor={\TheAuthors}
}
\fi

\def\arraystretch{0.75} 
\begin{document}
	\renewcommand\algorithmiccomment[1]{\hfill{\color{green!40!black}$\blacktriangleright$ #1}}
	
	\maketitle
	
	\begin{abstract}
	Bayesian regression remains a simple but effective tool based on Bayesian inference techniques. For large-scale applications, with complicated posterior distributions, Markov Chain Monte Carlo methods are applied. To improve the well-known computational burden of Markov Chain Monte Carlo approach for Bayesian regression, we developed a multilevel Gibbs sampler for Bayesian regression of linear mixed models. The level hierarchy of data matrices is created by clustering the features and/or samples of data matrices. Additionally, the use of correlated samples is investigated for variance reduction to improve the convergence of the Markov Chain. Testing on a diverse set of data sets, speed-up is achieved for almost all of them without significant loss in predictive performance. 
	\end{abstract}
	%
	%
	
\section{Introduction}
Linear regression remains a simple yet widely used technique in statistics and machine learning. Although more advanced techniques exist, linear models have the advantage that they offer a simple explanation of how the input variables influence the output variable(s). The simplest linear model is a linear combination between a weight vector $\textbf{b}\in\mathbb{R}^{F\times 1}$ and a data matrix $X\in\mathbb{R}^{N\times F}$ \begin{equation*}
\textbf{y}=X\textbf{b}
\end{equation*} with $\textbf{y}\in \mathbb{R}^{N\times 1}$ the target values, $N$ the number of observations and $F$ the number of features. This can be seen as statistical inference and the most common approaches are Bayesian inference and frequentist inference \cite{bishop2006pattern,gelman2013bayesian,friedman2001elements}.

Bayesian inference will draw any conclusions about the parameters or observations in terms of probability distributions. One advantage of the Bayesian approach is that it allows uncertainty before (prior) and after (posterior) the observations $X$ and $\textbf{y}$ have been incorporated. Using prior probability distributions allows the inclusion of information about the model parameters. Another advantage of the Bayesian approach is the ability to make probabilistic predictions for new data using the posterior of the parameter $\textbf{b}$.

For high dimensional spaces, using exact analytical techniques is not feasible or even impossible and the use of Monte Carlo (MC) methods is widely spread \cite{bishop2006pattern,gelman2013bayesian,robert2013monte,sorensen2007likelihood}. MC still requires a distribution to draw samples from a probability distribution and Markov Chain Monte Carlo (MCMC) is a general method to sample from arbitrary distributions. MCMC algorithms have given researchers the opportunity to investigate Bayesian inference for complicated statistical models. The most common MCMC methods are Metropolis-Hastings \cite{hastings1970monte,metropolis1953equation} and Gibbs sampling \cite{Gelfand1990Sampling,Geman1984Stochastic} with Gibbs sampling being a special case of Metropolis-hastings. 

Bayesian inference has been used for applications from genetics \cite{Coninck813,sorensen2007likelihood}, chemogenomics \cite{Simm2017Macau} and cell imaging \cite{Simm2018repurposing}. The data matrices in these fields are often large-scale and sparse. High performance computing (HPC) techniques are required for these data sets. More specifically, research has been developed to speed-up Metropolis-Hastings, since not not all the generated samples are accepted in the Markov Chain. In Metropolis-Hastings, computing the next sample in the chain is often computationally intensive and this sample is not always accepted in the Markov chain. Several techniques have been developed in different fields to improve the acceptance rate of Metropolis-Hastings, for example see \cite{Christen2005Markov, Efendiev2006PreconditioningMarkov,Hoang2013Complexity,Vandecasteele2020micromacro}. In contrast, Gibbs sampling does not have an acceptance step and Simm et al. \cite{Simm2017Macau} describe a Gibbs sampler based on solving a linear system using Krylov subspace methods and thus preserving the data sparsity essential for HPC.

We draw inspiration from HPC techniques that have been developed for large-scale partial differential equations (PDE) using multiple levels. The finite difference method approximates solution of the PDE using a spatial discretization step resulting in a linear system. By varying this discretization step, several levels of different accuracies are created as shown in Figure \ref{fig:grid_levels}. More specifically, discretizing the spatial variables results in a grid of nodes and the solution of the PDE is then approximated in these nodes. A fine discretization leads to many nodes. On the other hand, a coarse discretization results in less grid points. Both the fine and the coarse grid provide approximate solutions of the PDE, where the fine grid finds a better approximation to the exact solution but is more expensive to compute due to the larger number of grid points. By varying the discretization, a hierarchy of grids is created with increasing number of grid points. This approach is called Multigrid and exploits this hierarchy to create an efficient iterative solver or preconditioner for the underlying linear system of the PDE \cite{briggs2000multigrid,vassilevski2008multilevel}.  

While Multigrid exploits the hierarchy of grids when solving the resulting linear system, multilevel Monte Carlo (MLMC) for PDEs with random coefficients however exploits the same hierarchy of grids in the sampling process. As detailed in \cite{cliffe2011multilevel} most of the sampling variance is present in the coarsest grid and MLMC will take most of the samples on the coarsest and thus cheaper levels. MLMC has also been successfully applied for stochastic differential equations in \cite{doi:10.1287/opre.1070.0496}. More recently, Robbe et al \cite{Robbe2019Recycling} have combined Multigrid with MLMC for PDEs with random coefficients. The succes of MLMC for classical MC methods has led to increased interest in multilevel Metropolis-Hastings algorithms \cite{Dodwell2015Hierarchical}.            

\begin{figure*}[ht]
	\tikzsetnextfilename{lvlhierarchy}
	\begin{subfigure}{\textwidth}
		\centering
		\resizebox{\linewidth}{!}{\begin{tikzpicture}
\node at (0,-2.5) {Level 2};
\node at (5,-2.5) {Level 1};
\node at (10,-2.5) {Level 0};
\draw[step=0.5,lightgray] (-2,-2) grid (2,2);
 \foreach \x in {0,...,8}
\foreach \y in {0,...,8}
{
	\fill (\x/2-2,\y/2-2) circle (2.5pt);
}

\draw[step=1,lightgray] (3,-2) grid (7,2);
\foreach \x in {0,...,4}
\foreach \y in {0,...,4}
{
	\fill (\x+3,\y-2) circle (2.5pt);
}

\draw[step=2,lightgray] (8,-2) grid (12,2);
\foreach \x in {0,...,2}
\foreach \y in {0,...,2}
{
	\fill (2*\x+8,2*\y-2) circle (2.5pt);
}
\end{tikzpicture}}
		\caption{Hierarchy of Grids}
		\label{fig:grid_levels}
	\end{subfigure}
	\begin{subfigure}{\textwidth}
		\centering
		\resizebox{\linewidth}{!}{\begin{tikzpicture}
\node at (-1,-2.3) {$X^T_2$};
\node at (2.2,-0.3) {$X_2$};
\node at (5.5,-0.3) {$X^T_1$};
\node at (7.5,0.7) {$X_1$};
\node at (9.3,0.7) {$X^T_0$};
\node at (10.5,1.2) {$X_0$};
\node at (0.5,2.3) {Level 2};
\node at (6.5,2.3) {Level 1};
\node at (10,2.3) {Level 0};
\draw[step=0.5] (-2,-2) grid (0,2);
\draw[step=0.5] (0.499,0) grid (4.5,2);
\draw[step=0.5] (4.99,0) grid (6,2);
\draw[step=0.5] (6.499,0.99) grid (8.5,2);
\draw[step=0.5] (8.99,0.99) grid (9.5,2);
\draw[step=0.5] (9.99,1.499) grid (11,2);
\end{tikzpicture}}
		\caption{Hierarchy of dense data matrices for $X^TX$}
		\label{fig:mat_levels}
	\end{subfigure}
	\caption{Examples of the level hierarchies in PDEs and data. Figure \ref{fig:grid_levels} shows the hierarchy of grids for the discretization of a PDE. Figure \ref{fig:mat_levels} shows the hierarchy of levels for $X^tX$ using a data matrix $X$ and clustering both the rows and columns.}
	\label{fig:lvlhierarchy}
\end{figure*} 

We will design a multilevel version of the Gibbs sampler for Bayesian regression by Simm et al. \cite{Simm2017Macau}. In contrast with classical MLMC which varies the discretization step, a hierarchy of data matrices is created using clustering algorithms for the rows and or columns of the data matrix $X$. Figure \ref{fig:lvlhierarchy} shows an example for dense matrices with clustering for both rows and columns. Samples from these levels are then combined in one Markov Chain. Samples taken on the coarser levels are cheaper to compute, resulting in overall speed-up. 

The outline of the paper is as follows. First, we will provide further background on MCMC and MLMC in Section \ref{sect:prelim}. Next, the Gibbs sampler for Bayesian regression \cite{Simm2017Macau} is given and extended for linear mixed models in Section \ref{sect:noise_ibr}. This is followed by our multilevel Gibbs sampler in Section \ref{sect:mlmc-g} and our multilevel Monte Carlo Gibbs sampler in Section \ref{sect:correlation}. Finally results are given for a variety of data sets in Section \ref{sect:exp} and a conclusion is given in Section \ref{sect:conc}. 
\section{Preliminaries}\label{sect:prelim}
 Linear models are defined by \begin{equation}
\textbf{y}=X\textbf{b}+\textbf{e} \label{eq:lr}
\end{equation} with $\textbf{y}\in \mathbb{R}^{N\times 1}$ the $N$ sample observations, $X\in\mathbb{R}^{N\times F}$ a collected feature matrix, $F$ the number of features and $\textbf{e}\in\mathbb{R}^{N\times 1}$ the unknown residuals. Linear regression aims to find a distribution or point estimate for the weight vector $\textbf{b}\in\mathbb{R}^{F\times 1}$. 

Bayesian inference for linear models estimates the parameter $\textbf{b}$ using prior information and the Bayes rule. The Bayes rule is defined by\begin{equation*}
p(\textbf{b}|X)=\frac{p(X|\textbf{b})p(\textbf{b})}{p(X)}
\end{equation*}
with $p(\textbf{b}|X)$ the posterior, $p(X|\textbf{b})$ the likelihood, $p(\textbf{b})$ the prior and $p(X)$ the marginal likelihood. Using the posterior, probabilistic predictions can be made for new data $X_\text{new}$ with the predictive distribution \begin{equation*}
p(X_{\text{new}}|X)=\int p(X_{\text{new}}|\textbf{b})p(\textbf{b}|X)d\textbf{b}.
\end{equation*}  

Deriving the exact posterior distribution is not always possible. However, the quantity of interest is often the expected value of the posterior $p(\textbf{b}|X)$ or more generally of a function $f(\textbf{b})$ under a target distribution $p(\textbf{b})$ with $\textbf{b}$ a discrete or continuous variable. The expected value is defined by \begin{equation*}
F=\mathbb{E}[f(\textbf{b})]=\int f(\textbf{b})p(\textbf{b})d\textbf{b}.
\end{equation*} 
The Monte Carlo method (MC) approximates this expected value by drawing $H$ random and uncorrelated samples $\textbf{b}^{(h)}$ from $p(\textbf{b})$ and then takes the average
\begin{align}
\mathbb{E}[f(\textbf{b})]&\approx F^{MC}_{H} \\ &=\frac{1}{H}\sum_{h=1}^H f (\textbf{b}^{(h)}) \label{eq:mc}
\end{align} 
with $F^{MC}_{H}$ as the MC estimate for the expected value using $H$ samples. 

Equation (\ref{eq:mc}) estimates the expected value, but still requires to sample from the distribution $p(\textbf{b})$. Sampling from a specific distribution $p(\textbf{b})$ can be severely hard, particularly in high-dimensional spaces and $p(\textbf{b})$ might even be unavailable. Markov Chain Monte Carlo (MCMC) is widely used to sample from these distributions and is a general method for drawing samples from arbitrary posterior distributions \cite{bishop2006pattern,gelman2013bayesian}. The idea is to sample from proposal distributions and each sample is iteratively drawn from a distribution that approximates the target distribution better, with each sample used to improve and correct the proposal distribution. These samples are drawn sequentially, $\textbf{b}_1, \textbf{b}_2, \textbf{b}_3, \dots$, resulting in the Markov chain. Important is that each sample is thus drawn from a proposal distribution that increasingly better approximates the desired target distribution. A Markov Chain generally has a burn-in period where the first $H_{\text{burn\_in}}$ samples are not taken into account as these are not yet drawn from a distribution that is close enough to the target distribution.    

In order to create a Markov chain, a new sample $\textbf{b}^{(h)}$ is proposed using a transition distribution $T^h(\textbf{b}^{(h)},\textbf{b}^{(h-1)})$ with $\textbf{b}^{(h-1)}$ the previous sample. In a Markov chain, a distribution is called stationary or invariant if each new sample leaves that distribution unchanged. The target distribution $p^*(\textbf{b})$ should be invariant over the Markov chain and a sufficient condition is called detailed balance, defined as  \begin{equation*}
p^*(\textbf{b})T(\textbf{b},\textbf{b}')=p^*(\textbf{b}')T(\textbf{b}',\textbf{b})
\end{equation*} with $\textbf{b}$ and $\textbf{b}'$ two samples. The final to be mentioned property is ergodicity, which says that for any start distribution, the target distribution is invariant and for $h\to\infty$ the distribution $p(\textbf{b}^{(h)})$ converges to the target distribution $p^*(\textbf{b})$. The invariant distribution is then called the equilibrium distribution. 

The Metropolis-Hastings algorithm is widely used to generate samples in a Markov chain. For each sample $\textbf{b}^{(h)}$, a follow-up sample $\textbf{b}'$ is proposed from a jumping distribution $J_h(\textbf{b}',\textbf{b}^{(h)})$. The proposed sample is then accepted with probability \begin{equation*}
A(\textbf{b}',\textbf{b}^{(h)})=\min \left(1,\frac{p(\textbf{b}')J_h(\textbf{b}^{(h)},\textbf{b}')}{p(\textbf{b}^{(h)})J_h(\textbf{b}',\textbf{b}^{(h)})}\right).
\end{equation*}
If the sample is accepted, then $\textbf{b}^{(h+1)}=\textbf{b}'$ else $\textbf{b}^{(h+1)}=\textbf{b}^{(h)}$. Gibbs sampling can be considered as a special case of the more general Metropolis-Hastings algorithm \cite{bishop2006pattern,gelman2013bayesian}. Suppose the parameter $\textbf{b}$ consists of F subcomponents $b_i$ with $b_i$ the individual weight for linear regression. A Gibbs sampler will now sample each component $b_i$ individually assuming the other components $\textbf{b}_{\backslash i}$ are given using $p(b_i|\textbf{b}_{\backslash i})$. A Gibbs sampler has the advantage that each new sample is accepted. 

As detailed in \cite{Dodwell2015Hierarchical}, the mean square error (MSE) of the MCMC estimator is given by \begin{equation}
\epsilon(F_H^{MC})=\mathbb{E}_\textbf{B}\left[(F_H^{MC}-F)^2 \right] \label{eq:err1}
\end{equation}
with $\mathbb{E}_\textbf{B}$ the expected value with respect to the joint distribution of $\textbf{B}=\{\textbf{b}^{(h)}\}$. Equation (\ref{eq:err1}) is typically rewritten as
\begin{equation}
\epsilon(F_H^{MC})=\mathbb{V}_\textbf{B}\left[F_H^{MC}\right]+\left(\mathbb{E}_\textbf{B}\left[F_H^{MC}\right]-F\right)^2 \label{eq:err2}
\end{equation}  with $\mathbb{V}$ the variance. More details can be found in \cite{Dodwell2015Hierarchical,robert2013monte}. It is not possible to express the MSE in terms of $H$ due to the fact that the generated and consecutive samples in the Markov Chain are not independent. This is in contrast with classical Monte Carlo methods, where the MSE can be expressed in terms of the number of independent samples $H$, e.g. $\epsilon \sim\frac{1}{H}$. 

Classical Monte Carlo methods are used in applications for which the numerical solution depends on a discretization step. For these applications, a hierarchy of levels can be created. This is, for example, true for stochastic differential equations \cite{doi:10.1287/opre.1070.0496} using a time step or PDEs using a spatial discretization step \cite{cliffe2011multilevel} as seen in Figure \ref{fig:grid_levels}. Multilevel Monte Carlo (MLMC) now combines decreasing discretization steps resulting in several levels $l=0,\dots,L$  with $L$ the smallest discretization step and thus the finest level. By introducing this level hierarchy, the given notation is adjusted to incorporate the level parameter $l$. We denote with $f_l(\textbf{b})$ the function of interest on level $l$ and with $F_{l,H}^{MC}$ the MC estimate on level $l$ for $\mathbb{E}[f_l(\textbf{b})]$. Following the MLMC terminology, the telescoping sum uses this level hierarchy and exploits the linearity of the expectation operator \begin{equation}
\mathbb{E}[f_L(\textbf{b})]=\mathbb{E}[f_0(\textbf{b})]+\sum_{l=1}^L \mathbb{E}[f_{l}(\textbf{b})-f_{l-1}(\textbf{b})]. \label{eq:telesum}
\end{equation} 
Using Monte Carlo as unbiased estimator for the expectation combined with (\ref{eq:telesum}) results in \begin{equation*}
\mathbb{E}[f_L(\textbf{b})]\approx \frac{1}{H_0}\sum_{h=1}^{H_0}f_0(\textbf{b}^{(h)})+\sum_{l=1}^L \frac{1}{H_l}\sum_{h=1}^{H_l}\left(f_{l}(\textbf{b}^{(h)})-f_{l-1}(\textbf{b}^{(h)})\right)
\end{equation*}
 with $H_l$ the number of samples on level $l$. Denoting  \begin{equation}Y_{l,H_l}^{MC}=\frac{1}{H_l}\sum_{h=1}^{H_l}\left(f_{l}(\textbf{b}^{(h)})-f_{l-1}(\textbf{b}^{(h)})\right) \label{eq:mcdiff} \end{equation} results in
 \begin{align*}
 \mathbb{E}[f_L(\textbf{b})]&\approx F_{L,\{H_l\}}^{MLMC}\\&= F_{0,H_0}^{MC}+\sum_{l=1}^L Y_{l,H_l}^{MC}
 \end{align*}
 with $F_{L,\{H_l\}}^{MLMC}$ the MLMC estimate for $\mathbb{E}[f_L(\textbf{b})]$. 
  
  In short, there are two principles behind MLMC. The first principle is that taking samples on the coarser levels $l<L$ is computationally cheaper than taking samples on the finest level $L$. These cheaper samples however still approximate the same solution as the finest level with some loss in accuracy. The second principle can be seen as the use of a control variate. A control variate $E$ for a random variable $U$ is defined by the facts that taking samples from $E$ is cheap and $E$ is positively correlated with $U$. Exploiting the linearity of the expectation operator results in \begin{equation}
  \mathbb{E}[U]= \mathbb{E}[E]+ \mathbb{E}[U-E] \label{eq:cv}.
  \end{equation} Since $E$ is positively correlated with $U$, $\mathbb{V}[U-E]<\mathbb{V}[U]$ and an MC estimator for the right-hand side of \eqref{eq:cv} will require less samples taken from $U$ than a MC estimator for $\mathbb{E}[U]$ in order to reach the same MSE. Note the resemblance of \eqref{eq:cv} and the telescoping sum \eqref{eq:telesum}. In order for the coarser level in \eqref{eq:mcdiff} to act as a control variate, the samples $f_{l}(\textbf{b}^{(h)})$ and $f_{l-1}(\textbf{b}^{(h)})$ need to be positively correlated. It is thus important that when using MC for estimating the difference $\mathbb{E}[f_{l}(\textbf{b})-f_{l-1}(\textbf{b})]\approx \frac{1}{H_l}\sum_{h=1}^{H_l}(f_{l}(\textbf{b}^{(h)})-f_{l-1}(\textbf{b}^{(h)})$, that the $H_l$ pairs $\{f_{l}(\textbf{b}^{(h)}),f_{l-1}(\textbf{b}^{(h)})\}$ are uncorrelated but within a pair the samples $f_{l}(\textbf{b}^{(h)})$ and $f_{l-1}(\textbf{b}^{(h)})$ are positively correlated. The coarser level $l-1$ can then be seen as a control variate for the finer level $l$. In conclusion, the coarser levels in MLMC are constructed such that the coarser levels contain most of the variance and MLMC will as a result take more samples on the coarser levels and thus cheaper levels, resulting in less computational work \cite{cliffe2011multilevel,Dodwell2015Hierarchical,doi:10.1287/opre.1070.0496,Robbe2019Recycling}.   
   
Due to the success of MLMC in different fields, Multilevel Markov chain Monte Carlo has been investigated \cite{Dodwell2015Hierarchical, Vandecasteele2020micromacro}. By using a hierarchy of levels, the acceptance rate of Metropolis-Hastings can be improved. In \cite{Dodwell2015Hierarchical} they created a hierarchy of chains to improve the acceptance rate and additionally reduce the variance for subsurface flow. Currently, there is no generally accepted method to design Multilevel Markov Chain Monte Carlo methods. One problem is that the generated samples in MCMC are not independent as required by MLMC. One technique is to only consider each $k$-th sample in the chain and is called thinning. 
 
\section{Noise Injection Gibbs Sampler for Bayesian regression}\label{sect:noise_ibr}
In ordinary linear regression \cite{bishop2006pattern,gelman2013bayesian} the residuals are assumed uncorrelated and normally distributed in (\ref{eq:lr}), e.g. $\textbf{e}\sim \mathcal{N}(\textbf{0},\tau^{-1}I_N)$ with $I_N$ the identity matrix of size $\mathbb{R}^{N\times N}$ and $\tau > 0$ the precision parameter. In the Bayesian approach the observations $\textbf{y}$ are probabilistic \begin{equation*}
p(\textbf{y}|X,\textbf{b})=\prod_{i=1}^N \mathcal{N}(y_i|{\textbf{x}}_i{\boldsymbol{\textbf{b}}},\tau^{-1})
\end{equation*} with $\textbf{x}_i\in \mathbb{R}^{1\times F}$ the features of sample $i$. The conditional posterior distribution of $\textbf{b}$ given $\tau,X,\beta$ and $\textbf{y}$ is then \begin{equation*}
p(\textbf{b}|\tau,\textbf{y},X) \sim \mathcal{N}((X^TX+\beta I_F)^{-1}X^T\textbf{y},((X^TX+\beta I_F)\tau)^{-1})
\end{equation*}
with $\beta$ a regularization parameter. Computing the covariance matrix $((X^TX+\beta I_F)\tau)^{-1}$ is inconvenient and often computationally infeasible. Simm et al. \cite{Simm2017Macau} however describe a Gibbs sampler for Bayesian regression and using their noise injection trick, a sample of $\textbf{b}$ can be taken by solving \begin{equation*}
\Biggl({X}^T{X}+	\frac{\lambda_\textbf{b}}{\tau}I_F\Biggr){\boldsymbol{\textbf{b}}}= {X}^T(\textbf{y}+\textbf{e}_1)+\frac{\textbf{e}_2}{\tau}
\end{equation*}
with $\textbf{e}_1\sim \mathcal{N}(\textbf{0},\tau^{-1}I_N)$ and $\textbf{e}_2\sim \mathcal{N}(\textbf{0},\lambda_\textbf{b}I_F)$. A zero mean and uncorrelated normal distribution is used as prior for $\textbf{b}$: \begin{equation*}
p(\textbf{b},\lambda_\textbf{b}|\alpha_\textbf{b},\beta_\textbf{b})\sim \mathcal{N}(\textbf{b}|\textbf{0},\lambda_\textbf{b}^{-1}I_F)\mathcal{G}(\lambda_\textbf{b}|\alpha_\textbf{b},\beta_\textbf{b})
\end{equation*}
with $\mathcal{G}(\lambda_\textbf{b}|\alpha_\textbf{b},\beta_\textbf{b})$ the gamma distribution used as conjugate prior for the precision parameter $\lambda_\textbf{b}>0$. 
 
The noise injection trick can also be extended to the linear mixed model equations \cite{henderson1963selection,henderson1959estimation} \begin{equation}
\textbf{y}=W\textbf{v}+Z\textbf{u}+\textbf{e} \label{eq:mlm}
\end{equation}
with $\textbf{y}\in \mathbb{R}^{N\times 1}$ the observations and $N$ the sample size. The vector $\textbf{v}\in \mathbb{R}^{F\times 1}$ represents the $F$ fixed effects and $\textbf{u}\in \mathbb{R}^{S\times 1}$ are the $S$ random effects. The matrices $W\in\mathbb{R}^{N\times F}$ and $Z\in\mathbb{R}^{N\times S}$ are incidence matrices coupling the effects to the observations \cite{Coninck813}. We assume that the random effects and the residual error $\textbf{e}\in \mathbb{R}^{N\times 1}$ are normally distributed: $\textbf{u}\sim \mathcal{N}(\textbf{0},G)$ and $\textbf{e}\sim \mathcal{N}(\textbf{0},R)$. This formulation can further be simplified by assuming that the random effects are a priori uncorrelated, e.g. $G=\lambda_\textbf{u}^{-1}I_S$ with $\lambda_\textbf{u}>0$ the precision. As before, we assume that the residual errors ($\textbf{e}$) are uncorrelated, resulting in $R=\tau^{-1}I_N$ with $\tau>0$ the precision. 

The linear mixed model (\ref{eq:mlm}) can now be rewritten as \begin{equation*}
\textbf{y}=X\boldsymbol{\textbf{b}}+\textbf{e}
\end{equation*}
with $X=[W\ Z]$ and $\boldsymbol{\textbf{b}}=[\textbf{v}^T\ \textbf{u}^T]^T$, resulting in \begin{equation*}
p(\textbf{y}|X,\boldsymbol{\textbf{b}})=\prod_{i=1}^N \mathcal{N}(y_i|\textbf{x}_i\boldsymbol{\textbf{b}},\tau^{-1}). 
\end{equation*}
In the Bayesian setting \cite{gelman2013bayesian,sorensen2007likelihood}, the fixed effects are uncertain and assumed normally distributed $\boldsymbol{\textbf{v}}\sim \mathcal{N}(0|\lambda_\textbf{v}^{-1}I_F)$ with $\lambda_\textbf{v}>0$ the precision parameter. Using gamma distributions as priors for $\lambda_\textbf{v}$ and $\lambda_\textbf{u}$, the conditional probability is  
\begin{equation*}
p(\boldsymbol{\textbf{b}},\Lambda|\textbf{y},X)= \mathcal{N}(\boldsymbol{\textbf{b}}|0,\Lambda^{-1}) \mathcal{G}(\lambda_\textbf{v}|\alpha_\textbf{v},\beta_\textbf{v})\mathcal{G}(\lambda_\textbf{u}|\alpha_\textbf{u},\beta_\textbf{u})
\end{equation*} with \begin{equation}
\Lambda=\begin{bmatrix}
\lambda_\textbf{v}& & & & &0 \\
 &\ddots & & &\\
 & &\lambda_\textbf{v}& & & \\
& & &\lambda_\textbf{u} & & \\
& & & &\ddots & &\\
0& & & & &\lambda_\textbf{u} \\
\end{bmatrix}.\label{eqn:Lambdam}
\end{equation}	
Using the noise injection trick \cite{Simm2017Macau}, a sample of $\boldsymbol{\textbf{b}}$ can be taken by solving \begin{equation}
\Biggl(X^TX+	\frac{\Lambda}{\tau}\Biggr)\boldsymbol{\textbf{b}}= X^T(\textbf{y}+\textbf{e}_1)+\frac{\textbf{e}_2}{\tau} \label{eqn:noise_macau}
\end{equation} with $\textbf{e}_1=\mathcal{N}(\textbf{0},\tau^{-1}I)$ and $\textbf{e}_2=\mathcal{N}(\textbf{0},\Lambda)$. Samples for $\tau$, $\lambda_\textbf{v}$ and $\lambda_\textbf{u}$ are taken from respectively
\begin{align}
p(\tau|\boldsymbol{\textbf{b}},\textbf{y},X,\alpha_\textbf{e},\beta_\textbf{e})&=\mathcal{G}\biggl(\tau\biggl|\alpha_\textbf{e}+\frac{N}{2}, \beta_\textbf{e}+\sum_{i=1}^N\frac{(\textbf{y}_i-X_i\boldsymbol{\textbf{b}})^2}{2}\biggr) ,\label{eq:posttau} \\
p(\lambda_\textbf{v}|\boldsymbol{\textbf{b}},\textbf{y},X,\alpha_\textbf{v},\beta_\textbf{v})&=\mathcal{G}\biggl(\lambda_\textbf{v}\biggl|\alpha_\textbf{v}+\frac{F}{2}, \beta_\textbf{v}+\sum_{f=1}^{F}\frac{\boldsymbol{\textbf{b}}_f^2}{2}\biggr),\label{eq:postlb}\\
p(\lambda_\textbf{u}|\boldsymbol{\textbf{b}},\textbf{y},X,\alpha_\textbf{u},\beta_\textbf{u})&=\mathcal{G}\biggl(\lambda_\textbf{u}\biggl|\alpha_\textbf{u}+\frac{S}{2}, \beta_\textbf{u}+\sum_{s=F+1}^{F+S}\frac{\boldsymbol{\textbf{b}}_s^2}{2}\biggr).\label{eq:postlu}
\end{align}
The noise injection Gibbs sampler for linear mixed models is given in Algorithm \ref{alg:nilmm}. 
\begin{algorithm}[H]
\caption{Noise injection for linear mixed models}
\label{alg:nilmm}
\begin{algorithmic}[1]
	\STATE \textbf{Input}: \begin{itemize}
		\item matrix X
		\item hyperparameters $\alpha_\textbf{e}$, $\alpha_\textbf{v}$, $\alpha_\textbf{u}$, $\beta_\textbf{e}$, $\beta_\textbf{v}$ and $\beta_\textbf{u}$
	\end{itemize} 
	\STATE sample $\tau$, $\lambda_\textbf{v}$ and $\lambda_\textbf{u}$ from priors
	\STATE generate  $\textbf{e}_1=\mathcal{N}(\textbf{0},\tau^{-1}I)$ and $\textbf{e}_2=\mathcal{N}(\textbf{0},\Lambda)$ using (\ref{eqn:Lambdam})
	\STATE sample $\textbf{b}^{(1)}$ by solving \ref{eqn:noise_macau}
	\FOR{$h=2,\dots, H$}
	\STATE sample $\tau$, $\lambda_\textbf{v}$ and $\lambda_\textbf{u}$ from (\ref{eq:posttau}), (\ref{eq:postlb}) and (\ref{eq:postlu})
	\STATE generate  $\textbf{e}_1=\mathcal{N}(\textbf{0},\tau^{-1}I)$ and $\textbf{e}_2=\mathcal{N}(\textbf{0},\Lambda)$ using (\ref{eqn:Lambdam})
	\STATE sample $\textbf{b}^{(h)}$ by solving \eqref{eqn:noise_macau}
	\ENDFOR
	\RETURN $\mathbb{E}[\textbf{y}]\approx \frac{1}{H}\sum_{h=1}^H X\textbf{b}^{(h)}$
\end{algorithmic}
\end{algorithm} 

\section{Hierarchical Markov Chain Monte-Carlo for Bayesian regression}\label{sect:mlmc-g}
We are interested in $\mathbb{E}[\textbf{y}]=\mathbb{E}[X\boldsymbol{b}]$. Using the Gibbs sampler approach, we approximate \begin{equation*}
\mathbb{E}[\textbf{y}]\approx \frac{1}{H}\sum_{h=1}^H X\textbf{b}^{(h)}
\end{equation*} with $\textbf{b}^{(h)}$ the samples taken by solving (\ref{eqn:noise_macau}). For large-scale data, solving thousands linear systems (\ref{eqn:noise_macau}) can be computationally expensive even when iterative solvers are used. Therefore, we want to create a hierarchy of levels approximating the data set $X$ resulting in data levels with increasing computational work and data information. We propose to use clustering algorithms to create this hierarchy of $L+1$ data matrices $X_l$ for $l=0\dots L$ by clustering the features (columns) of $X=X_L$. This strategy has been applied to create a two-level preconditioner for Bayesian regression \cite{DBLP:journals/corr/abs-1806-05826}. A coarser data matrix is created by defining the coarse features as $$\textbf{c}_\mathcal{S}=1/\sqrt{|\mathcal{S}|}\sum_{i\in\mathcal{S}} X(:,i)$$ with $\textbf{c}_\mathcal{S}$ the coarse feature representing the features of cluster $\mathcal{S}$ and $|S|$ the number of features in $\mathcal{S}$. When the features in $X_L=[X_1, X_2, \dots, X_{F_C}]$ are ordered per cluster with ${F_C}$ the number of clusters and $X_i$ the data matrix with the features of cluster $i$, we can define a coarser data matrix as $X_{L-1}=X_LP_L$  with 		\begin{equation}
P_L^T=\begin{array}{cccccccccccc}	
&\multicolumn{3}{c}{\overbrace{\hspace{60pt}}^{n_1}}&\multicolumn{3}{c}{\overbrace{\hspace{60pt}}^{n_2}}&\dots &\multicolumn{3}{c}{\overbrace{\hspace{65pt}}^{n_{F_C}}}& \\
\ldelim[{5.3}{10pt}[] &\frac{1}{\sqrt{n_1}} &\ldots &\frac{1}{\sqrt{n_1}} & & & & & & & &\rdelim]{5.3}{10pt}[]  \\
&  & & &\frac{1}{\sqrt{n_2}} &\ldots  &\frac{1}{\sqrt{n_2}} &  & & & &  \\
&  & & & & & & \ddots& & & & \\
& & & & & & & &\frac{1}{\sqrt{n_{F_C}}} &\ldots &\frac{1}{\sqrt{n_{F_C}}} & \\
\end{array} \label{eq:prolong}.
\end{equation}  
Note that this definition of $P_L$ leads to \begin{eqnarray*}
	P_L^T(X_L^TX_L+\beta I)P_L &=	P_L^TX_L^TX_LP_L+P_L^T\beta IP_L\\
	&= X_{L-1}^TX_{L-1}+\beta \underbrace{P_L^TP_L}_{=I}
\end{eqnarray*} as detailed in \cite{DBLP:journals/corr/abs-1806-05826}. 

The linear system \eqref{eqn:noise_macau} involves $X^TX$ and this matrix-product is used in principle component analysis (PCA). PCA finds the eigenvectors with the largest eigenvalues of $X^TX$ and is a projection method that maximizes total data variance \cite{bishop2006pattern,DBLP:journals/corr/abs-1806-05826}. By clustering the features, ideally the variance within a cluster is as small as possible. This means that the coarser levels can approximate the larger eigenvectors and contain most of the data variance. Note that applying PCA directly to create the coarser level is not feasible since it requires the eigenvalue decomposition of $X^TX$.    

Using a hierarchy of data matrices, it is possible to define a Gibbs sampler on each level. One simple approach is to define a Markov Chain for each level. This would result in $L+1$ Markov Chains and thus $L+1$ burn-ins. It is, however, possible to interpolate a sample $\textbf{b}_{l}$ on level $l$ to level $l+1$ with $\textbf{b}_{l+1}=P_{l+1}\textbf{b}_{l}$ for $l=0\dots L-1$. This interpolated sample is then used to sample $\tau$, $\lambda_\textbf{v}$ and $\lambda_\textbf{u}$ in equations (\ref{eq:posttau}), (\ref{eq:postlb}) and (\ref{eq:postlu}). Next, a sample $\textbf{b}_{l+1}$ on level $l+1$  is drawn using equation (\ref{eqn:noise_macau}). This provides us with samples for the parameter $\textbf{b}_{l}$ for each level $l$. The quantity of interest is the expected value of the predicted observations $\mathbb{E}[\textbf{y}]$. Since $X_l P_{l}= X_{l-1}$ and thus $X_{l-1}\textbf{b}_{l-1}=X_l P_{l}\textbf{b}_{l-1}$. 

This means that $\mathbb{E}[\textbf{y}]$ can be estimated using the fine level data matrix $X_L$ and coarser solutions $\textbf{b}_l$ with $$\mathbb{E}[\textbf{y}]\approx X_L\frac{\prod_{k=l}^{L-1}P_{k+1}\textbf{b}_l}{H_l}$$ for $l=0, \dots,L$. This is important since generally we are interested in the predictions of new and unseen data $X_{\text{test}}$ with the model trained using different data $X_{\text{train}}$. 
Combining the samples from each level $l$, $\mathbb{E}[\textbf{y}]$ is approximated by \begin{equation}
\mathbb{E}[\textbf{y}]\approx X_L\frac{\sum_{l=0}^{L}\left(\prod_{k=l}^{L-1}P_{k+1}\left(\sum_{h=1}^{H_l}\textbf{b}_l^{(h)}\right)\right)}{\sum_{l=0}^{L} H_l}.\nonumber
\end{equation} Multilevel Gibbs sampling using noise injection for linear mixed models is given in Algorithm \ref{alg:MLMCMC}. 
 
\begin{algorithm}[H]
	\caption{Multilevel Noise injection Gibbs sampler for linear mixed models}
	\label{alg:MLMCMC}
	\begin{algorithmic}[1]
	\STATE \textbf{Input}: \begin{itemize}
	\item Hierarchy of matrices $X_l$ for $l=0,\dots,L$
	\item hyperparameters $\alpha_\textbf{e}$, $\alpha_\textbf{v}$, $\alpha_\textbf{u}$, $\beta_\textbf{e}$, $\beta_\textbf{v}$ and $\beta_\textbf{u}$
\end{itemize} 
\STATE Take $H_0$ samples $\textbf{b}_0^{(h)}$ on the \textbf{coarsest level} $l=0$ using the Noise Injection Algorithm \ref{alg:nilmm}
\STATE $\boldsymbol{\Sigma_0}= X_L\prod_{k=0}^{L-1}P_{k+1}\left(\sum_{h=1}^{H_0} \textbf{b}_0^{(h)}\right)$
\STATE Sample on the \textbf{remaining levels}
\begin{ALC@g}
\FOR{$l=1,\dots, L$}
\STATE interpolate $\textbf{b}_{l}=P_l\textbf{b}_{l-1}^{(H_{l-1})}$
\FOR{$h=1,\dots, H_l$}
\STATE sample $\tau$, $\lambda_\textbf{v}$ and $\lambda_\textbf{u}$ from (\ref{eq:posttau}), (\ref{eq:postlb}) and (\ref{eq:postlu})
\STATE generate  $\textbf{e}_1=\mathcal{N}(\textbf{0},\tau^{-1}I)$ and $\textbf{e}_2=\mathcal{N}(\textbf{0},\Lambda)$ using (\ref{eqn:Lambdam})
\STATE sample $\textbf{b}^{(h)}$ by solving \eqref{eqn:noise_macau} using $X_l$
\ENDFOR
\STATE $\boldsymbol{\Sigma_l}= X_L\prod_{k=l}^{L-1}P_{k+1}\left(\sum_{h=1}^{H_l} \textbf{b}_l^{(h)}\right)$
\ENDFOR
\end{ALC@g}
\RETURN $\mathbb{E}[\textbf{y}]\approx \frac{\sum_{l=0}^{L}\boldsymbol{\Sigma_l}}{\sum_{l=0}^L H_l}$
\end{algorithmic}
\end{algorithm}

\section{Multilevel Markov Chain Monte Carlo using correlated samples}\label{sect:correlation}
MLMC has proven to speed-up Monte Carlo methods for different applications. As explained in Section \ref{sect:mlmc:mlmc}, positive correlation between samples on different levels results in variance reduction of the MC estimator. Before detailing the correlation for the Gibbs sampler, the telescoping sum \eqref{eq:telesumpde} is first applied to linear models\begin{align}
\mathbb{E}[\textbf{y}_L]&=\mathbb{E}[\textbf{y}_0]+\sum_{l=1}^{L}\mathbb{E}[\textbf{y}_l-\textbf{y}_{l-1}] \label{eq:lmm_mlmc}\\
&=\mathbb{E}[X_0\textbf{b}_0]+\sum_{l=1}^{L}\mathbb{E}[X_l\textbf{b}_l-X_{l-1}\textbf{b}_{l-1}]\nonumber\\
&=X_0\mathbb{E}[\textbf{b}_0]+\sum_{l=1}^{L}X_{l}\mathbb{E}[\textbf{b}_l-P_{l}\textbf{b}_{l-1}] \nonumber\\
&\approx X_0\sum_{h=1}^{H_0} \textbf{b}_0^{(h)}+\sum_{l=1}^{L}X_{l}\frac{1}{H_l}\sum_{h=1}^{H_l}\textbf{b}_l^{(h)}-P_{l}\textbf{b}_{l-1}^{(h)} \label{eq:tele_gibbs_pbl}  
\end{align}
and consists of the expected value on the coarsest level and $L$ correcting terms. For the MLMC scheme to be profitable, the samples $\textbf{b}_l^{(h)}$ and $P_l\textbf{b}_{l-1}^{(h)}$ in (\ref{eq:tele_gibbs_pbl}) should be correlated. Two approaches to ensure correlation between the samples are considered. The first approach is simply to define $ P_{l}^T\textbf{b}_{l}=\textbf{b}_{l-1}$ and this approach thus takes approximately the average within a cluster (up to a factor) of the fine solution as the coarse value which results in \begin{equation}\mathbb{E}[\textbf{y}_l-\textbf{y}_{l-1}]\approx X_L\prod_{k=l}^{L-1}P_{k+1}\left(\frac{1}{H_l}\sum_{h=1}^{H_l} \textbf{b}_l^{(h)}-P_lP_l^T\textbf{b}_{l}^{(h)}\right). \label{eq:projections}
\end{equation} This is a simple projection of the finer solution to the coarser solution. It is, however, important that the finer samples $\textbf{b}_l^{(h_l)}$ of the difference on level $l=i$ are drawn from the same posterior as the coarser samples $\textbf{b}_{l-1}^{(h_{l+1})}$ for the next level $l=i+1$ for $i=0,\dots, L-1$. Simply aggregating the finer solution does not guarantee this. 

The second approach solves (\ref{eqn:noise_macau}) to draw sample $\textbf{b}_{l-1}$ on the coarser level with exactly the same values $\tau,\lambda_\textbf{v}$, $\lambda_\textbf{u}$ and $\textbf{e}_1$ used for drawing sample $\textbf{b}^{l}$ on the finer level. Only the random vector $\textbf{e}_2$ is projected to the coarser space $P_l^T\textbf{e}_{2,l}=\textbf{e}_{2,l-1}$. Empirical results will show that the second approach is computationally more expensive but obtains better results. The algorithm using the second approach for the correlated samples is outlined in Algorithm \ref{alg:MLMCMLMC}.
\begin{algorithm}[H]
	\caption{Multilevel Monte Carlo noise injection Gibbs sampler using linear systems solves}
	\label{alg:MLMCMLMC}
	\begin{algorithmic}[1]
			\STATE \textbf{Input}: \begin{itemize}
			\item Hierarchy of matrices $X_l$ for $l=0,\dots,L$
			\item hyperparameters $\alpha_\textbf{e}$, $\alpha_\textbf{v}$, $\alpha_\textbf{u}$, $\beta_\textbf{e}$, $\beta_\textbf{v}$ and $\beta_\textbf{u}$
		\end{itemize} 
		\STATE Take $H_0$ samples $\textbf{b}_0^{(h)}$ on the \textbf{coarsest level} $l=0$ using the Noise Injection Algorithm \ref{alg:nilmm}
		\STATE $\mathbb{E}[\textbf{y}_0]\approx X_L\prod_{k=0}^{L-1}P_{k+1}\left(\frac{1}{H_0}\sum_{h=1}^{H_0} \textbf{b}_0^{h}\right)$
		\STATE Sample on the \textbf{remaining levels}
		\begin{ALC@g}
			\FOR{$l=1,\dots, L$}
			\STATE interpolate $\textbf{b}_{l}=P_l\textbf{b}_{l-1}^{(H_{l-1})}$
			\FOR{$h=1,\dots, H_l$}
			\STATE sample $\tau$, $\lambda_\textbf{v}$ and $\lambda_\textbf{u}$ from (\ref{eq:posttau}), (\ref{eq:postlb}) and (\ref{eq:postlu})
			\STATE generate  $\textbf{e}_1=\mathcal{N}(\textbf{0},\tau^{-1}I)$ and $\textbf{e}_2=\mathcal{N}(\textbf{0},\Lambda)$ using (\ref{eqn:Lambdam})
			\STATE sample $\textbf{b}^{(h)}_{l}$ by solving \eqref{eqn:noise_macau} using $X_l$
			\STATE restrict $\textbf{e}_{2,l-1}=P_l^T\textbf{e}_{2,l}$
			\STATE Using $X_{l-1}$ and $\textbf{e}_{2,l-1}$ sample $\textbf{b}^{(h)}_{l-1}$ by solving \ref{eqn:noise_macau} reusing $\tau$, $\lambda_\textbf{v}$, $\lambda_\textbf{u}$ and $\textbf{e}_1$ 	
			\STATE store difference $\textbf{d}_l^{(h)}=\textbf{b}_l^{(h)}-P_l\textbf{b}_{l-1}^{(h)}$
			\ENDFOR
			\STATE $\mathbb{E}[\textbf{y}_l-\textbf{y}_{l-1}]\approx X_L\prod_{k=l}^{L-1}P_{k+1}\left(\frac{1}{H_l}\sum_{h=1}^{H_l} \textbf{d}_l^{(h)}\right)$
			\ENDFOR
			\RETURN  $\mathbb{E}[\textbf{y}]\approx \mathbb{E}[\textbf{y}_0]+\sum_{l=1}^{L}\mathbb{E}[\textbf{y}_l-\textbf{y}_{l-1}]$	
		\end{ALC@g}
	\end{algorithmic}
\end{algorithm}  

\begin{figure*}[ht]
	\begin{subfigure}{0.43\textwidth}
		\tikzsetnextfilename{vcycle}
		\centering
		\resizebox{\linewidth}{!}{\begin{tikzpicture}
\node[left of=abel=level 2] at (0,0) {\tiny level 2};
\node[left of=abel=level 1] at (0,-0.5) {\tiny level 1};
\node[left of=abel=level 0] at (0,-1) {\tiny level 0};
\node[circle, fill, minimum size=3.5pt,inner sep=0pt, outer sep=0pt] at (-0.5,0) {};
\node   [circle, fill, minimum size=3.5pt,inner sep=0pt, outer sep=0pt] at (-0.35,0.0) {};
\node (a2) [circle, fill, minimum size=3.5pt,inner sep=0pt, outer sep=0pt] at (-0.2,0.0) {};

\node(c1)[circle, fill, minimum size=3.5pt,inner sep=0pt, outer sep=0pt] at (-0.1,-0.5) {};
\node   [circle, fill, minimum size=3.5pt,inner sep=0pt, outer sep=0pt] at (0.05,-0.5) {};
\node (c2)[circle, fill, minimum size=3.5pt,inner sep=0pt, outer sep=0pt] at (0.2,-0.5) {};

\node (e1)[circle, fill, minimum size=3.5pt,inner sep=0pt, outer sep=0pt] at (0.3,-1) {};
\node   [circle, fill, minimum size=3.5pt,inner sep=0pt, outer sep=0pt] at (0.45,-1) {};
\node  (e2)[circle, fill, minimum size=3.5pt,inner sep=0pt, outer sep=0pt] at (0.6,-1) {};

\node (d1) [circle, fill, minimum size=3.5pt,inner sep=0pt, outer sep=0pt] at (0.7,-0.5) {};
\node  (d) [circle, fill, minimum size=3.5pt,inner sep=0pt, outer sep=0pt] at (0.85,-0.5) {};
\node (d2) [circle, fill, minimum size=3.5pt,inner sep=0pt, outer sep=0pt] at (1.0,-0.5) {};

\node (b1)[circle, fill, minimum size=3.5pt,inner sep=0pt, outer sep=0pt] at (1.1,0) {};
\node  [circle, fill, minimum size=3.5pt,inner sep=0pt, outer sep=0pt] at (1.25,0.0) {};
\node[circle, fill, minimum size=3.5pt,inner sep=0pt, outer sep=0pt] at (1.4,0.0) {};
\draw [-] (a2) -- (c1);
\draw [-] (c2) -- (e1);
\draw [-] (e2) -- (d1);
\draw [-] (d2) -- (b1);
\end{tikzpicture}}
		\caption{three level V-cycle}
		\label{fig:Vcycle}
	\end{subfigure}
	\begin{subfigure}{0.55\textwidth}
		\tikzsetnextfilename{wcycle}
		\centering
		\resizebox{\linewidth}{!}{\begin{tikzpicture}
\node[left of=abel=level 2] at (0,0) {\tiny level 2};
\node[left of=abel=level 1] at (0,-0.5) {\tiny level 1};
\node[left of=abel=level 0] at (0,-1) {\tiny level 0};
\node[circle, fill, minimum size=3.5pt,inner sep=0pt, outer sep=0pt] at (-0.5,0) {};
\node   [circle, fill, minimum size=3.5pt,inner sep=0pt, outer sep=0pt] at (-0.35,0.0) {};
\node (a2) [circle, fill, minimum size=3.5pt,inner sep=0pt, outer sep=0pt] at (-0.2,0.0) {};

\node(c1)[circle, fill, minimum size=3.5pt,inner sep=0pt, outer sep=0pt] at (-0.1,-0.5) {};
\node   [circle, fill, minimum size=3.5pt,inner sep=0pt, outer sep=0pt] at (0.05,-0.5) {};
\node (c2)[circle, fill, minimum size=3.5pt,inner sep=0pt, outer sep=0pt] at (0.2,-0.5) {};

\node (e1)[circle, fill, minimum size=3.5pt,inner sep=0pt, outer sep=0pt] at (0.3,-1) {};
\node   [circle, fill, minimum size=3.5pt,inner sep=0pt, outer sep=0pt] at (0.45,-1) {};
\node  (e2)[circle, fill, minimum size=3.5pt,inner sep=0pt, outer sep=0pt] at (0.6,-1) {};

\node (d1) [circle, fill, minimum size=3.5pt,inner sep=0pt, outer sep=0pt] at (0.7,-0.5) {};
\node  (d) [circle, fill, minimum size=3.5pt,inner sep=0pt, outer sep=0pt] at (0.85,-0.5) {};
\node (d2) [circle, fill, minimum size=3.5pt,inner sep=0pt, outer sep=0pt] at (1.0,-0.5) {};

\node (f1)[circle, fill, minimum size=3.5pt,inner sep=0pt, outer sep=0pt] at (1.1,-1) {};
\node   [circle, fill, minimum size=3.5pt,inner sep=0pt, outer sep=0pt] at (1.25,-1) {};
\node  (f2)[circle, fill, minimum size=3.5pt,inner sep=0pt, outer sep=0pt] at (1.4,-1) {};

\node (g1) [circle, fill, minimum size=3.5pt,inner sep=0pt, outer sep=0pt] at (1.5,-0.5) {};
\node  (d) [circle, fill, minimum size=3.5pt,inner sep=0pt, outer sep=0pt] at (1.65,-0.5) {};
\node (g2) [circle, fill, minimum size=3.5pt,inner sep=0pt, outer sep=0pt] at (1.8,-0.5) {};

\node (b1)[circle, fill, minimum size=3.5pt,inner sep=0pt, outer sep=0pt] at (1.9,0) {};
\node  [circle, fill, minimum size=3.5pt,inner sep=0pt, outer sep=0pt] at (2.05,0.0) {};
\node[circle, fill, minimum size=3.5pt,inner sep=0pt, outer sep=0pt] at (2.2,0.0) {};
\draw [-] (a2) -- (c1);
\draw [-] (c2) -- (e1);
\draw [-] (e2) -- (d1);
\draw [-] (d2) -- (f1);
\draw [-] (f2) -- (g1);
\draw [-] (g2) -- (b1);
\end{tikzpicture}}
		\caption{three level W-cycle}
		\label{fig:Wcycle}
	\end{subfigure}
	\caption{Three level V-cycle and W-cycle with three samples after each change of level.}
	\label{fig:VWcycle}
\end{figure*}
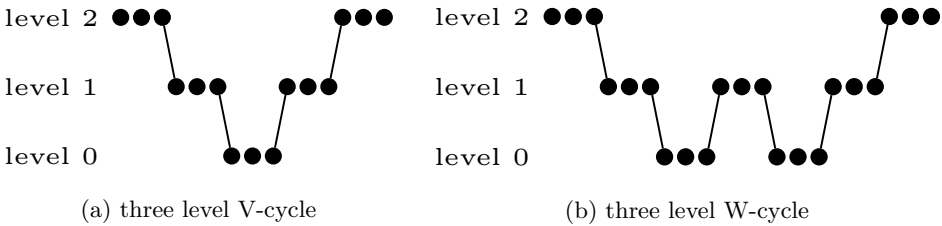

Algorithms \ref{alg:MLMCMC} and \ref{alg:MLMCMLMC} start sampling at the coarsest level and then refine until the finest level. This means that all the samples on each level are drawn consecutively. In contrast to multilevel Monte Carlo, the samples of a Markov Chain are not {i.i.d.} and consecutive samples are correlated. Instead of taking all the samples at once for each level, it is possible to take a few samples at one level, then move to another level and take a few samples before changing again to another level. Figure \ref{fig:VWcycle} shows examples of traversing the levels. Figure \ref{fig:Vcycle} details a V-cycle with three samples per level and figure \ref{fig:Wcycle} a W-cycle. Restricting the previous sample $\textbf{b}_l$ from a finer level to a coarser level $\textbf{b}_{l-1}$ is done using $\textbf{b}_{l-1}=P^T_l\textbf{b}_{l}$.

\section{Experiments}\label{sect:exp}
 We have investigated the described multilevel sampling schemes for Bayesian regression. Using real data $X$ but simulated $\textbf{b}$ and $\textbf{y}$, the model was trained with $\textbf{y}_{\text{train}}=X_\text{train}\textbf{b}+\textbf{e}$ and evaluated using $\textbf{y}_{\text{test}}=X_\text{test}\textbf{b}$ with $X_\text{train}$ and $X_\text{test}$ determined by $5$-fold cross-validation. The characteristics of the different data matrices $X$ are given in Table~\ref{tab:data}. Most are available online as indicated. Generating normal distributed $\textbf{b}\sim \mathcal{N}(0,10I_F)$ and $\textbf{e}\sim \mathcal{N}(0,10^3I_N)$, $\textbf{y}_{\text{train}}$  was simulated. 
 		
The application of the ChEMBl data is chemogenomics. They predict the activity of chemical compounds on certain proteins. More precisely, one of the indicators for drug-protein activity is the IC50 value. IC50 or the half maximal inhibitory concentration measures the substance concentration required to inhibit the activity of a protein by $50\%$. The bioactivity database ChEMBL version 19 \cite{bento2014chembl} is publicly available. These chemical compounds are modeled by combinations of molecules. These chemical substructures are described by the extended-connectivity fingerprints (ECFP \cite{rogers2010extended}) and were computed using rdkit \cite{rdkit} with 3 layers. More details can be found in \cite{TAVERNIER201986}. 

The single-nucleotide polymorphism (SNP) data sets are used in genomic prediction for animals or plants \cite{Coninck813}. This application uses a linear mixed model \eqref{eq:mlm}. The matrix $Z$ was simulated using AlphaMPSim \cite{AlphaMPSim} and the random effects $\textbf{u}$ represent the SNP marker effects. The number of SNP markers were set to $10\ 000$, $50\ 000$ and $100\ 000$ resulting in three different data sets. Only the overall mean was modeled for the fixed effects resulting in $W=\textbf{1}$. For the SNP data sets, the observations $\textbf{y}$ were available and used.

 \begin{table}[ht]
 	\centering
 	\scalebox{0.82}{
 		\begin{tabular}{ lp{3cm}rrrrr} \toprule
 			Name&Source&value type &\#rows&\#columns&\#nonzeros&fillin($\%$) \\ \midrule
 			Trec10&\cite{davis2011university}&double &106&478&8612&17.00\\
 			CNAE&\cite{Lichman2013}&double &1080&856&7233&0.78\\							
 			micromass&\cite{Lichman2013}&double &360&1300&48\ 713&10.41\\
 			DrivFace&\cite{Lichman2013}&double &606&6400&3\ 878\ 400&100.00\\
 			arcene&\cite{Lichman2013}&double &100&10\ 000&540\ 941&54.09\\
 			SNP1&\cite{AlphaMPSim}&short int &10\ 000&10\ 000&57\ 641\ 064&57.64\\
 			tmc2007& SIAM 2007 Text Mining competition&double &21\ 519&30\ 438&2\ 283\ 179&0.35\\
 			nlpdata&\cite{MatlabOTB}&double &31\ 572&34\ 023&2\ 277\ 757&0.21\\				
 			rcv1\_multi&\cite{Lewis:2004:RNB:1005332.1005345}&double &15\ 564&47\ 236&1\ 028\ 284&0.14\\
 			SNP2&\cite{AlphaMPSim} &short int&10\ 000&50\ 000&282\ 212\ 533&56.44\\
 			news20&\cite{keerthi2005modified}&double &15\ 935&62\ 061&1\ 272\ 569&0.13\\
 			SNP3&\cite{AlphaMPSim}&short int &10\ 000&100\ 000&570\ 261\ 477&57.03\\
 			E2006&\cite{kogan-etal-2009-predicting}&double &16\ 087&150\ 360&19\ 971\ 015&0.83\\
 			ChEMBL&\cite{bento2014chembl}&binary &167\ 668&291\ 714&12\ 246\ 376&0.03	\\					
 			\bottomrule
 	\end{tabular}}
 	\caption{ The characteristics of the data matrices $X$ used in the experiments. }\label{tab:data}
 \end{table}
 
Leader-follower clustering \cite{hartigan1975clustering} was recursively used to cluster features, resulting in a hierarchy of data matrices. The performance of clustering algorithmns is often data dependent and Leader-follower clustering has the advantage that the algorithm only passes trough the data once as detailed in \cite{DBLP:journals/corr/abs-1806-05826}. Table~\ref{tab:data_lvls} shows the number of levels and the corresponding feature sizes on the different levels for the used data sets. Leader-Follower clustering uses a threshold and passes trough the data and assigns data points to a cluster if the point is within the threshold. For the data sets with $10\ 000$ are more features the coarse size was defined in the range $[3000,8500]$ since computing the singular value decomposition is feasible in this range. The threshold can be selected by starting with a large threshold, for which leader-follower clustering is computed fast,  and then increased until the coarse size is within the defined range.        
 
 \begin{table}[ht]
 	\centering
 	\scalebox{0.8}{
 		\begin{tabular}{ l|c|l}
 			Name& Number of levels & Feature sizes\\ \midrule		
 			Trec10&3&[181, 319, 478]\\
			CNAE&3&[198, 354, 856]\\							
			micromass&3&[542, 754, 1300]\\
			DrivFace&3&[2061, 3907, 6400]\\
			arcene&3&[3341, 5933, 10000]\\
			SNP1&3&[3582, 5178, 10000]\\
			tmc2007&3&[5881, 18709, 30438]\\
			nlpdata&3&[5258, 12095, 34023]\\				
			rcv1\_multi&3&[6858, 18363, 47236]\\
			SNP2&4&[7647, 15764, 28326, 50000]\\
			news20&4&[5922, 17811, 31000, 62061]\\
			SNP3&4&[8235, 21608, 44928, 100000]\\
			E2006&4&[7647, 28326, 78091, 150360]\\
			ChEMBL&6&[5664, 17019, 50382, 112612, 170223, 291714]	\\	
 			\bottomrule
 	\end{tabular}}
 	\caption{The number of levels and average feature sizes for the different data sets used in the experiments.}
 	 	\label{tab:data_lvls}%
 \end{table}

 \begin{table}[ht]
	\centering
	\scalebox{1.0}{
		\begin{tabular}{ l|p{9.5cm}} \toprule
			Abbreviation&Description\\\midrule
			Gibbs&Noise injection Gibbs sampler given in Algorithm \ref{alg:nilmm}\\
			ML-G&Noise injection Multilevel (ML) Gibbs sampler (G) given in Algorithm \ref{alg:MLMCMC}\\
			MLCSS-G&Noise injection Multilevel (ML) Gibbs sampler (G)  with correlated samples (CS) using linear solves (S) given in Algorithm \ref{alg:MLMCMLMC}\\
			MLCSP-G&Noise injection Multilevel (ML) Gibbs sampler (G) with correlated samples (CS) using projections (P) as detailed in \eqref{eq:projections}\\ 
     		MLMLP-G&Noise injection Multilevel (ML) Multilevel-preconditioned (MLP) Gibbs sampler (G) given in Algorithm \ref{alg:MLMCMC}\\
			MLMLPCSS-G&Noise injection Multilevel (ML) Multilevel-preconditioned (MLP) Gibbs sampler (G) with correlated samples (CS) using linear solves (S) given in Algorithm \ref{alg:MLMCMLMC}\\					
			MLMLPCSP-G&Noise injection Multilevel (ML) Multilevel-preconditioned (MLP) Gibbs sampler (G) with correlated samples (CS) using projections (P) as detailed in \eqref{eq:projections}\\
			W-cycle(30)& W-cycle level sample configuration with 30 samples at each level before changing to the next level. \\
			\bottomrule
	\end{tabular}}
	\caption{Abbreviations.}\label{tab:abbreviations}
\end{table}
The level hierarchy created by clustering can be used as a two-level preconditioner to accelerate the convergence of CG \cite{DBLP:journals/corr/abs-1806-05826}. This generally results in lower execution times for solving the linear system in the sampling process. This does, however, requires the computation of the singular value decomposition on the coarsest level. We investigated both CG as solver on each level and CG accelerated by a two-level preconditioner with two iterations of CG as pre-smoothing for the multilevel sampling \cite{DBLP:journals/corr/abs-1806-05826}. Table~\ref{tab:abbreviations} shows the abbreviations of all the possible different samplers in the experiments.

Note that sampling distributions of $\tau$, $\lambda\textbf{v}$, $\lambda\textbf{u}$ in equations \eqref{eq:posttau}, \eqref{eq:postlb} and \eqref{eq:postlu} are determined by the solution $\textbf{b}$ found by solving the linear system. As a result, the sampling process is influenced by the underlying solver. Note that CG has a natural regularization role determined by the required tolerance or maximum number of iterations. Solving exactly on the coarse level can lead to different solutions for better or worse. Especially when using the correlated samples with system solves (MLCSS or MLMLPCSS) on the adjacent levels, the solvers on both levels should be of the same nature, e.g. iterative and thus not exact. If this not the case, the solutions on both levels can be too different, resulting in noise for the difference $\textbf{b}_l^{(h)}-P_l\textbf{b}_{l-1}^{(h)}]$ in \eqref{eq:tele_gibbs_pbl}. In the presented experiments, CG was always used on the coarsest level for fair comparison with all the other methods. Note that using an exact solver on the coarsest level for multilevel sampling without correlated samples is possible and would lead to even faster execution time with slightly different solutions. 

Table~\ref{tab:overview} provides the average setup time, execution time, speed-up, pearson correlation $\rho$, root mean square error (RMSE) and mean absolute error (MAE) for the given data sets using the noise injection Gibbs sampler (Gibbs), the multilevel Gibbs sampler (ML-G), the multilevel preconditioned multilevel Gibbs sampler (MLMLP-G) and the multilevel preconditioned multilevel Gibbs sampler with correlated samples using systems solves (MLMLPCSS-G) on the test set. 2200 Samples were taken with 200 discarded as burn-in. Non-informative priors were used setting $\alpha_\textbf{e}=1.0$, $\beta_\textbf{e}=1.0$, $\alpha_\textbf{v}=1.0$, $\beta_\textbf{v}=10^{-3}$, $\alpha_\textbf{u}=1.0$. The burn-in samples of the multilevel samplers were taken at the coarsest level and the remaining samples were taken consecutively and equally distributed on the different levels starting at the coarsest level. The experiments were implemented in C++\footnote{\url{https://scm.cs.kuleuven.be/scm/git/multilevel_macau}} and compiled with gcc 9.3.0 and OPENMP 4.0 with compile option -O3. The experiments were run on a machine with an Intel(R) Core(TM) i7-6560U (2.20GHz) processor with an L3 cache memory of 4096 KB and 16 GB of DRAM where 3 out of 4 cores were used. The reported values for the different experiments are the average of 5-fold cross-validation. 

As can be seen from Table~\ref{tab:overview} taking all the samples on the fine level (Gibbs) often results in the slightly better predictions. Using the multilevel sampling techniques result in small loss of accuracy, but for almost all data sets within range of the standard deviation. The finest level does contain all of the information, so it is natural that this leads to the best predictions. The fine level does, however, contain more noise components which are less present in the coarser level and the multilevel approach can achieve better accuracy as is the case for the SNP3 data set. Note that for the SNP3 data set, there is a significant improvement using the multilevel sampling schemes.

\begin{landscape}
	\begin{table}[ht]
		\centering
	\renewcommand{\arraystretch}{1.15}
		\begin{tabular}{cc}
		\scalebox{0.6}{
		\begin{tabular}{ lr|llll}
	& &Gibbs& ML-G & MLMLP-G & MLMLPCSS-G   \\ \midrule			
			& setup (s)  & &1.70E-2 &2.48E-2 &2.40E-2  \\
			& exec. (s)  &4.99 &4.44 &\textbf{2.92} & 5.89 \\
			Trec10 & speed-up  &1.0 &1.12&	\textbf{1.71}&	0.85\\
			& $\rho$  &\textbf{0.765} (0.106) &0.755 (0.107)&0.755 (0.107)&0.697  (0.159)\\
			& RMSE  &\textbf{6.44E2} (2.98E2) &6.55E2 (3.00E2)&6.55E2 (3.01E2) & 7.24E2 (2.57E2)\\
			& MAE  &\textbf{4.06E2} (1.53E2) &4.19E2 (1.54E2)&4.19E2 (1.54E2)& 4.68E2 (1.31E2)\\ \midrule	

			& setup (s)  & &2.84E-2 &1.07E-1 &1.18E-1 \\
			& exec. (s)  &8.96E-1 & 7.88E-1 &5.22 &8.87\\
			 CNAE & speed-up  &1.0&  \textbf{1.14}&	0.17&	0.10 \\
			& $\rho$  &\textbf{0.647} (0.035)&0.644 (0.034)&0.644 (0.034)&0.643  (0.036)\\
			& RMSE  &\textbf{6.59} (4.63E-1)&6.60 (4.42E-1)&6.60 (4.42E-1)& 6.62 (4.37E-1)\\
			& MAE  &4.97 (3.37E-1)&4.97 (3.50E-1)&4.97 (3.50E-1)& 4.98 (3.54E-1)\\ \midrule	

			& setup (s)  & &9.59E-2 &1.61E-1 &1.75E-1  \\
			& exec. (s)  &4.64E1 &4.22E1 &1.56E1 &2.66E1  \\
			micromass & speed-up  &1.0& 1.10&	\textbf{2.97}&	1.74\\
			& $\rho$  &0.984 (0.001)&0.984 (0.001)&0.984 (0.001)&0.984 (0.001) \\
			& RMSE  &\textbf{3.33E7} (3.93E5)&3.35E7 (4.16E5)&3.41E7 (5.49E5) &3.41E7  (5.41E5)\\
			& MAE  &\textbf{2.18E7} (3.16E5)&2.21E7 (3.34E5)&2.26E7 (3.80E5)& 2.27E7 (3.80E5)\\ \midrule	

			& setup (s)  & & 1.14E1 & 1.17E1& 1.16E1 \\
			& exec. (s)  &1.13E3 &6.69E2 &2.40E2 &3.88E2  \\
			DrivFace & speed-up  &1.0 & 1.69&\textbf{4.71}&	2.91 \\
			& $\rho$  &0.974 (0.005)&0.975 (0.005)&0.975 (0.005)&0.974  (0.005)\\
			& RMSE  &9.41 (1.49)&\textbf{9.36} (1.54)&\textbf{9.36} (1.5)& 9.40 (1.58)\\
			& MAE  &\textbf{6.99} (9.55E-1)&7.00 (1.00)&7.00 (1.0)& \textbf{6.99} (1.07)\\ \midrule	
	
			& setup (s)  & &1.95E2 &1.18E3 & 1.15E3 \\
			& exec. (s)  &6.18E3 &5.86E3 &3.24E3 &5.29E3  \\
			SNP1 & speed-up  &1.0 &1.05 &\textbf{1.91} &1.17  \\
			& $\rho$  &0.958 (0.003)&0.958 (0.003)&0.958 (0.003)& 0.958 (0.003)\\
			& RMSE  &1.47E-1 (4.16E-3)&1.47E-1 (4.20E-3)&1.47E-1 (4.19E-3)& 1.47E-1 (3.84E-3)\\
			& MAE  &1.16E-1 (3.48E-3)&1.16E-1 (3.56E3)&1.16E-1 (3.55E-3)& 1.16E-1 (3.26E-3)\\ \midrule	
			
			& setup (s)  & &8.00 &1.50E3 &1.49E3  \\
			& exec. (s)  &2.43E3 &1.30E3 &4.95E2&7.59E2  \\
			tmc2007 & speed-up  &1.0&1.87& \textbf{4.91}& 	3.20\\
			& $\rho$  &\textbf{0.988} (0.000)&0.984 (0.001)&0.982 (0.001)&0.985 (0.001) \\
			& RMSE  &\textbf{4.21E-1} (3.36E-3)&4.97E-1 (8.99E-3)&5.32E-1 (7.47E-3)&4.79E-2  (4.91E-3)\\
			& MAE  &\textbf{3.06E-1} (3.02E-3)&3.71E-1 (7.25E-3)&4.00E-1 (5.89E-3)&3.57E-1  (4.00E-3)\\
			\bottomrule
		\end{tabular}}
		
		&
			\scalebox{0.6}{
		\begin{tabular}{ lr|llll}

				& &Gibbs& ML-G & MLMLP-G & MLMLPCSS-G   \\ \midrule			
				& setup (s)  & &7.09 &1.91E3 &1.93E3  \\
				& exec. (s)  &3.23E2 &2.71E2 &1.98E2 &3.35E2  \\
				nlpdata & speed-up  &1.0& 1.19&	\textbf{1.63}&	0.96\\
				& $\rho$  &0.955 (0.014) &0.954 (0.015)&0.955 (0.015)&0.955  (0.014)\\
				& RMSE  &\textbf{7.09E1} (6.15)&7.24E1 (8.62)&7.20E1 (8.19) &7.12E1  (6.92)\\
				& MAE  &\textbf{3.59E1} (8.56E-1)&3.60E1 (9.16E-1)&3.60E1 (8.95E-1) &\textbf{3.59E1} (8.68E-1)  \\ \midrule	
					
				& setup (s)  & &5.51 &3.25E3 &3.24E3  \\
				& exec. (s)  &1.11E3 &5.60E2 &1.39E3 &1.98E3  \\
				rcv\_multi & speed-up  &1.0&\textbf{1.98} &	0.80& 0.56\\
				& $\rho$  &\textbf{0.893} (0.002)&0.883 (0.002)&0.882 (0.002)&0.838 (0.004) \\
				& RMSE  &\textbf{1.35} (1.59E-2)&1.41 (1.99E-2)&1.42 (1.95E-2)&1.64  (1.25E-2)\\
				& MAE  &\textbf{8.96E-1} (4.9E-3)&9.75E-1 (1.34E-2)&9.83E-1 (1.37E-2)&1.17  (1.09E-2)\\ \midrule	
			
				& setup (s)  & &1.53E3 &2.99E3 & 2.97E3 \\
				& exec. (s)  &4.34E4 &1.86E4 &8.91E3 &1.41E3  \\
				SNP2 & speed-up  &1.0 &2.33 &\textbf{4.87} &3.07  \\
				& $\rho$  &0.959 (0.002)&\textbf{0.961} (0.002)&\textbf{0.961} (0.002)& 0.960 (0.002)\\
				& RMSE  &1.57E-1 (2.00E-3)&\textbf{1.52E-1} (2.15E-3)&\textbf{1.52E-1} (2.16E-3)&1.54E-1  (2.22E-3)\\
				& MAE  &1.24E-1 (1.95E-3)&\textbf{1.20E-1} (2.00E-3)&\textbf{1.20E-1} (2.01E-3)&1.22E-1  (1.90E-3)\\ \midrule	
				
				& setup (s)  & &9.83 &1.65E3 &1.65E3  \\
				& exec. (s)  &7.22E3 &3.59E3 &4.22E3 &6.75E3  \\
				news20 & speed-up  &1.0  &\textbf{2.01}&1.71&	1.07\\
				& $\rho$  &\textbf{0.912} (0.017)&0.899 (0.015)&0.899 (0.015)& 0.892 (0.007)\\
				& RMSE  &\textbf{2.87E1} (8.39)&3.10E1 (1.05E1)&3.10E1 (1.05E1)& 3.17E1 (9.14)\\
				& MAE  &\textbf{1.27E1} (5.50E-1)&1.33E1 (5.95E-1)&1.33E1 (5.95E-1)&  1.43E1 (5.31E-1)\\ \midrule	
			
				& setup (s)  & &1.61E4 &2.20E4 & 2.14E4 \\
				& exec. (s)  &1.38E5 &3.85E4 &1.78E4 &2.78E4  \\
				SNP3 & speed-up  &1.0&	3.58&\textbf{7.75}&	4.96 \\
				& $\rho$  &0.955 (0.002)&\textbf{0.962} (0.002)&\textbf{0.962} (0.002)& 0.961 (0.002)\\
				& RMSE  &1.99E-1 (3.51E-3)&\textbf{1.79E-1} (5.14E-3)&\textbf{1.79E-1} (5.15E-3)&1.84E-1  (4.94E-3)\\
				& MAE  &1.57E-1 (3.55E-3)&\textbf{1.41E-1} (4.84E-3)&\textbf{1.41E-1} (4.84E-3) &1.45E-1  (4.88E-3)\\ \midrule	

				& setup (s)  & &2.02E2 &2.76E3  &2.75E3  \\
				& exec. (s)  &7.76E3 &4.12E3 &1.77E3&2.85E3  \\
				E2006 & speed-up  &1.0& 1.88&\textbf{4.38}&2.72\\
				& $\rho$  &\textbf{0.962} (0.002)&0.959 (0.002)&0.959 (0.002)&0.961 (0.002) \\
				& RMSE  &\textbf{7.88E-2} (2.56E-3)&8.11E-2 (2.76E-3)&8.21E-2 (2.53E-3)& 8.01E-2 (2.59E-3)\\
				& MAE  &\textbf{4.96E-2} (8.91E-3)&5.160E-2 (8.02E-4)&5.27E-2 (8.18E-4)& 5.10E-2 (1.31E-3) \\ \midrule

				& setup (s)  & &4.24E2 &4.63E3 &4.74E3  \\
				& exec. (s)  &1.72E3 &1.99E3 &1.35E3 &2.44E3 \\
				ChEMBL & speed-up  &1.0 & 0.86& \textbf{1.27}&	0.70\\
				& $\rho$  &\textbf{0.772} (0.005)&0.766 (0.004)&0.766 (0.004)&0.768  (0.004)\\
				& RMSE  &\textbf{1.54E1} (1.14E-1)&1.57E1 (9.32E-2)&1.57E1 (9.58E-2)& 1.56E1 (1.05E-1)\\
				& MAE  &\textbf{1.22E1} (8.05E-2)&1.23E1 (6.38E-2)&1.24E1 (6.53E-2)& 1.23E1 (7.47E-2)\\
			\bottomrule
		\end{tabular}%
		}
		\end{tabular}%
		\caption{Average setup time (setup), execution time (exec.), pearson correlation ($\rho$), root-mean-square error (RMSE) and mean absolute error (MAE) in scientific notation for the given data sets using Gibbs, ML-G, MLMLP-G and MLMLPCSS-G. The standard deviation for the accuracy measures is given within round brackets.}
		\label{tab:overview}%
	\end{table}%
\end{landscape}

The speed-up achieved by the multilevel schemes varies for the different data sets. Part of the speed-up depends on the use of preconditioning. Using the preconditioner results in larger setup times due to the calculation of the SVD but generally results in faster execution times since the linear systems are solved more efficiently. For the ChEMBL data the multilevel sampling without preconditioning is slower than taking all the samples on the fine level due to the fact that the nonzeros are binary on the fine level and floating point numbers on the coarser levels. The multilevel sampling with correlated samples results in better accuracy for some data sets but not consistently. The execution time for the correlated samples with linear systems (MLMLPCSS-G) has increased since, for each sample, a linear system has to be solved on two adjacent levels.
 
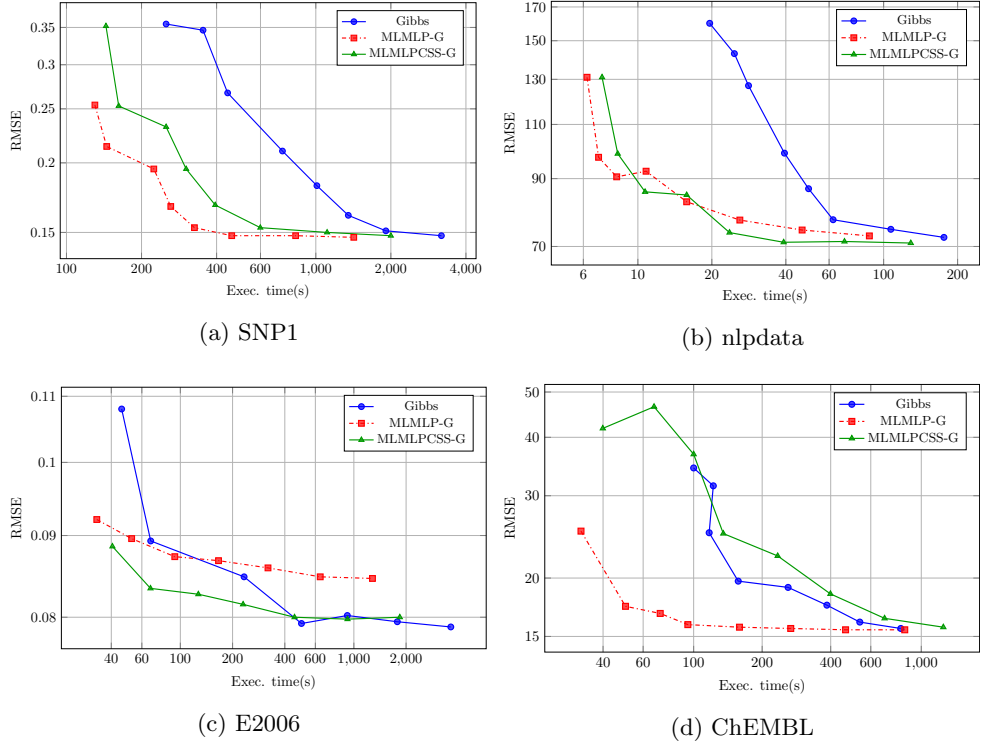
\begin{figure}
	\centering
	\begin{subfigure}{.495\textwidth}
		\tikzsetnextfilename{snp1is}%
		\resizebox{\linewidth}{!}{\begin{tikzpicture}
\begin{axis}[
height=8cm,
width=12cm,
xmode = log,
cycle list name=clusters,
ymode = log,
grid=both,
grid style={line width=.1pt, draw=gray!20},
major grid style={line width=.2pt,draw=gray!60},
minor tick num=5,
 log ticks with fixed point,
ytick={0.15,0.20,0.25,0.30,0.35},
xtick={20,40,60,100,200,400,600,1000,2000,4000},
xlabel=Exec. time(s),ylabel=RMSE,
legend columns=1,
legend pos=north east,
legend style={font=\small}
]
\addplot+ table[x=time_cg, y=rmse_cg] {Tikz/SNP1_iter.dat};
\addplot+ table[x=time_ml, y=rmse_ml] {Tikz/SNP1_iter.dat};
\addplot+ table[x=time_mlmcs, y=rmse_mlmcs] {Tikz/SNP1_iter.dat};
\legend{Gibbs, MLMLP-G,MLMLPCSS-G }
\end{axis}
\end{tikzpicture} }
		\caption{SNP1}
		\label{fig:snp1_is_rmse}
	\end{subfigure}
	\begin{subfigure}{.495\textwidth}
		\tikzsetnextfilename{nlpdatais}%
		\resizebox{\linewidth}{!}{\begin{tikzpicture}
\begin{axis}[
height=8cm,
width=12cm,
xmode = log,
cycle list name=clusters,
ymode = log,
grid=both,
grid style={line width=.1pt, draw=gray!20},
major grid style={line width=.2pt,draw=gray!60},
minor tick num=5,
 log ticks with fixed point,
ytick={70,90,110,130,150,170},
xtick={2,4,6,10,20,40,60,100,200},
xlabel=Exec. time(s),ylabel=RMSE,
legend columns=1,
legend pos=north east, 
legend style={font=\small}
]
\addplot+ table[x=time_cg, y=rmse_cg] {Tikz/nlpdata_iter.dat};
\addplot+ table[x=time_ml, y=rmse_ml] {Tikz/nlpdata_iter.dat};
\addplot+ table[x=time_mlmcs, y=rmse_mlmcs] {Tikz/nlpdata_iter.dat};
\legend{Gibbs, MLMLP-G,MLMLPCSS-G }
\end{axis}
\end{tikzpicture} }
		\caption{nlpdata}
		\label{fig:nlpdata_is_rmse}
	\end{subfigure}
	\begin{subfigure}{.495\textwidth}
		\tikzsetnextfilename{E2006is}%
		\resizebox{\linewidth}{!}{\begin{tikzpicture}
\begin{axis}[
height=8cm,
width=12cm,
xmode = log,
cycle list name=clusters,
ymode = log,
grid=both,
grid style={line width=.1pt, draw=gray!20},
major grid style={line width=.2pt,draw=gray!60},
minor tick num=5,
 log ticks with fixed point,
ytick={0.08,0.09,0.1,0.11 },
xtick={20,40,60,100,200,400,600,1000,2000},
xlabel=Exec. time(s),ylabel=RMSE,
legend columns=1,
legend pos=north east, 
legend style={font=\small}
]
\addplot+ table[x=time_cg, y=rmse_cg] {Tikz/E2006_iter.dat};
\addplot+ table[x=time_ml, y=rmse_ml] {Tikz/E2006_iter.dat};
\addplot+ table[x=time_mlmcs, y=rmse_mlmcs] {Tikz/E2006_iter.dat};
\legend{Gibbs, MLMLP-G,MLMLPCSS-G }
\end{axis}
\end{tikzpicture} }
		\caption{E2006}
		\label{fig:E2006_is_rmse}
	\end{subfigure}
	\begin{subfigure}{.495\textwidth}
		\tikzsetnextfilename{ChEMBLis}%
		\resizebox{\linewidth}{!}{\begin{tikzpicture}
\begin{axis}[
height=8cm,
width=12cm,
xmode = log,
cycle list name=clusters,
ymode = log,
grid=both,
grid style={line width=.1pt, draw=gray!20},
major grid style={line width=.2pt,draw=gray!60},
minor tick num=10,
 log ticks with fixed point,
ytick={15,20,30,40,50,60},
xtick={20,40,60,100,200,400,600,1000,2000},
xlabel=Exec. time(s),ylabel=RMSE ,
legend columns=1,
legend pos=north east, 
legend style={font=\small}
]
\addplot+ table[x=time_cg, y=rmse_cg] {Tikz/ChEMBL_iter.dat};
\addplot+ table[x=time_ml, y=rmse_ml] {Tikz/ChEMBL_iter.dat};
\addplot+ table[x=time_mlmcs, y=rmse_mlmcs] {Tikz/ChEMBL_iter.dat};
\legend{Gibbs, MLMLP-G,MLMLPCSS-G }
\end{axis}
\end{tikzpicture} }
		\caption{ChEMBL}
		\label{fig:ChEMBL_is_rmse}
	\end{subfigure}
	\caption{The average RMSE in function of the execution time in seconds for SNP1, nlpdata, E2006 and ChEMBL for Gibbs sampling, MLMLP-G and MLMLPCSS-G.}
	\label{fig:increasing_samples}
\end{figure}

Since taking samples at the coarser levels is cheaper, more samples can be taken at the coarse level in the same execution time as at the finest level. Table~\ref{tab:increasing_samples} shows the number of samples taken in different stages for $3$ and $4$ levels. Figure~\ref{fig:increasing_samples} shows the RMSE for four data sets in function of the execution time using the stages in Table~\ref{tab:increasing_samples}.
\begin{table}[ht]
	\centering
	\scalebox{0.8}{
		\begin{tabular}{ c|l|l|l}
			Stage& [$H_0$, $H_1$, $H_2$] & [$H_0$, $H_1$, $H_2$, $H_3$] & $H$\\ \midrule		
			1 & [2, 2, 2]       & [2, 2, 2, 2]			& 2 \\
			2 & [8, 4, 2]       & [16, 8, 4, 2]			& 4\\
			3 & [16, 8, 4]      & [32, 16, 8, 4]			& 16\\
			4 & [32, 16, 8]     & [64, 32, 16, 8]		& 64\\
			5 & [64, 32, 16]    & [128, 64, 32, 16]		& 128\\
			6 & [128, 64, 32]   & [256, 128, 64, 32] 	& 256\\
			7 & [256, 128, 64]  & [512, 256, 128, 64] 	& 512\\
			8 &	[512, 256, 128] &	[1024, 512, 256, 128] 	& 1024\\
			\bottomrule
	\end{tabular}}
	\caption{The increasing number of samples $H_l$ for the different levels }
	\label{tab:increasing_samples}%
\end{table}
  
Figure \ref{fig:increasing_samples} shows that the multilevel sampler (MLMLP-G) achieves better accuracy faster. Additionally the multilevel sampler converges faster than only sampling at the fine level (Gibbs) in Figures \ref{fig:snp1_is_rmse}, \ref{fig:nlpdata_is_rmse} and \ref{fig:ChEMBL_is_rmse}. For the E2006 data set the convergence is fast for all samplers. For the ChEMBL data, only the 4 coarsest levels were used and no samples were taken at the two finest levels. The multilevel sampler with correlated samples (MLMLPCSS-G) results in slightly better accuracy for two data sets than without correlated samples but requires more execution time. 
  
\subsection{Samples configuration}
For the results in Table \ref{tab:overview}, the samples are taken consecutively on each level. One can easily change levels while building the chain and Table \ref{tab:config_samples} shows the number of samples taken at each level for different configurations using $3$ and $6$ levels. 

\begin{table}[ht]
	\centering
	\scalebox{0.8}{
		\begin{tabular}{ c|c|c}
			Config& 3 levels & 6 levels \\\midrule	
			V-cycle(3) &[498, 999, 501]&[201, 399, 399, 399, 399, 201]\\
			V-cycle(10) &[500, 1000, 500]&[200, 400, 400, 400, 400, 200]\\
			V-cycle(30) &[480, 990, 510]&[210, 390, 390, 390, 390, 210]\\
			V-cycle(100)&[500, 1000, 500]&[200, 400, 400, 400, 400, 200]\\
			W-cycle(3) &[666, 999, 333]&[522, 783, 387, 189, 90, 27]\\
			W-cycle(10)&[670, 1000, 330]&[530, 790, 390, 190, 80, 20]\\
			W-cycle(30) &[660, 990, 330]&[570, 840, 390, 150, 30, 0]\\
			W-cycle(100)&[700, 1000, 300]&[600, 900, 400, 100, 0, 0]\\	
			\bottomrule
	\end{tabular}}
	\caption{The number of samples $H_l$ given as [$H_0, \dots, H_L $] for the different configurations used in the experiments.}
		\label{tab:config_samples}%
\end{table}

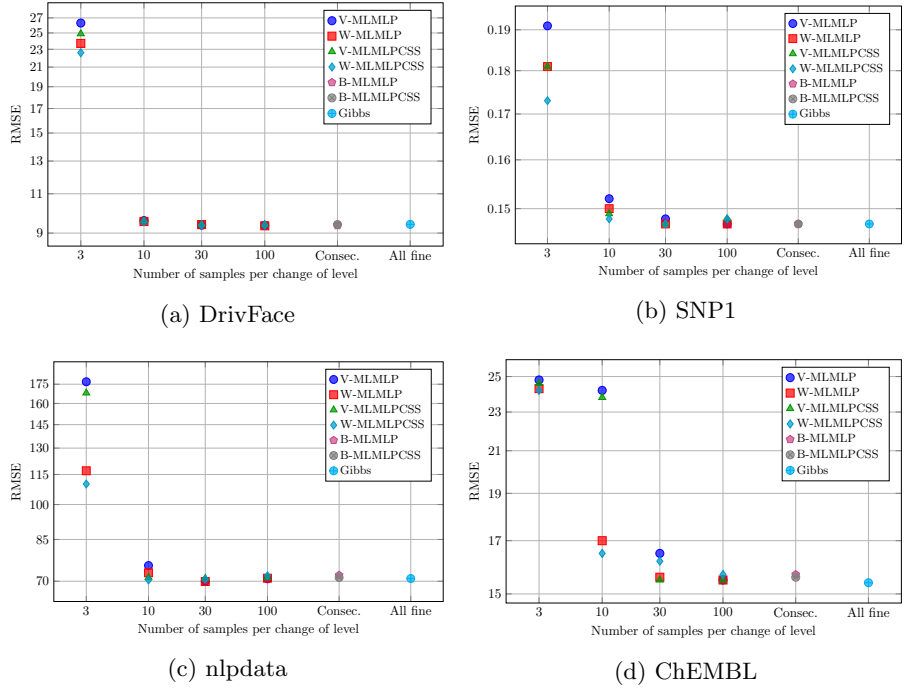
\begin{figure*}[ht]
	\centering
	\tikzsetnextfilename{DrivFace_vwcycle}
	\begin{subfigure}{0.45\textwidth}
		\centering
		\resizebox{\linewidth}{!}{\begin{tikzpicture}
\begin{axis}[
height=8cm,
width=12cm,
xmode = log,
cycle list name=clusters,
ymode = log,
grid=both,
grid style={line width=.1pt, draw=gray!20},
major grid style={line width=.2pt,draw=gray!60},
minor tick num=5,
 log ticks with fixed point,
 ytick={9,11,13,15,17,19,21,23,25,27},
xtick={3,10,30,100,400,1600},
xticklabels={3,10,30,100,Consec.,All fine},
xlabel=Number of samples per change of level,ylabel=RMSE,
legend pos=north east,
legend cell align={left},
legend style={font=\small}
]
\addplot+[only marks,mark size=3pt] table[x=ns, y=vrmse] {Tikz/DrivFace_vw.dat};
\addplot+[only marks,mark size=3pt ] table[x=ns, y=wrmse] {Tikz/DrivFace_vw.dat};
\addplot+[only marks,mark size=3pt] table[x=ns, y=vmcsrmse] {Tikz/DrivFace_vw.dat};
\addplot+[only marks,mark size=3pt ] table[x=ns, y=wmcsrmse] {Tikz/DrivFace_vw.dat};
\addplot+[only marks,mark size=3pt ] coordinates {
	(400, 9.36)
};
\addplot+[only marks,mark size=3pt ] coordinates {
	(400,  9.4)
};
\addplot+[only marks,mark size=3pt ] coordinates {
	(1600,  9.41)
};
\legend{V-MLMLP, W-MLMLP,V-MLMLPCSS, W-MLMLPCSS, B-MLMLP, B-MLMLPCSS, Gibbs}
\end{axis}
\end{tikzpicture}}
		\caption{DrivFace}
		\label{fig:vwcycle_DrivFace}
	\end{subfigure}
	\tikzsetnextfilename{snp1_vwcycle}
	\begin{subfigure}{0.45\textwidth}
		\centering
		\resizebox{\linewidth}{!}{\begin{tikzpicture}
\begin{axis}[
height=8cm,
width=12cm,
xmode = log,
cycle list name=clusters,
ymode = log,
grid=both,
grid style={line width=.1pt, draw=gray!20},
major grid style={line width=.2pt,draw=gray!60},
minor tick num=5,
 log ticks with fixed point,
 ytick={0.15,0.16,0.17,0.18,0.19,0.2},
xtick={3,10,30,100,400,1600},
xticklabels={3,10,30,100,Consec.,All fine},
xlabel=Number of samples per change of level,ylabel=RMSE,
legend pos=north east,
legend cell align={left},
legend style={font=\small}
]
\addplot+[only marks,mark size=3pt] table[x=ns, y=vrmse] {Tikz/snp1_vw.dat};
\addplot+[only marks,mark size=3pt ] table[x=ns, y=wrmse] {Tikz/snp1_vw.dat};
\addplot+[only marks,mark size=3pt] table[x=ns, y=vmcsrmse] {Tikz/snp1_vw.dat};
\addplot+[only marks,mark size=3pt ] table[x=ns, y=wmcsrmse] {Tikz/snp1_vw.dat};
\addplot+[only marks,mark size=3pt ] coordinates {
	(400, 0.147)
};
\addplot+[only marks,mark size=3pt ] coordinates {
	(400,  0.147)
};
\addplot+[only marks,mark size=3pt ] coordinates {
	(1600,  0.147)
};
\legend{V-MLMLP, W-MLMLP,V-MLMLPCSS, W-MLMLPCSS, B-MLMLP, B-MLMLPCSS, Gibbs}
\end{axis}
\end{tikzpicture}}
		\caption{SNP1}
		\label{fig:vwcycle_snp1}
	\end{subfigure}
		\tikzsetnextfilename{nlpdata_vwcycle}
	\begin{subfigure}{0.45\textwidth}
		\centering
		\resizebox{\linewidth}{!}{\begin{tikzpicture}
\begin{axis}[
height=8cm,
width=12cm,
xmode = log,
cycle list name=clusters,
ymode = log,
grid=both,
grid style={line width=.1pt, draw=gray!20},
major grid style={line width=.2pt,draw=gray!60},
minor tick num=5,
 log ticks with fixed point,
 ytick={70,85,100,115,130,145,160,175},
xtick={3,10,30,100,400,1600},
xticklabels={3,10,30,100,Consec.,All fine},
xlabel=Number of samples per change of level,ylabel=RMSE,
legend pos=north east,
legend cell align={left},
legend style={font=\small}
]
\addplot+[only marks,mark size=3pt] table[x=ns, y=vrmse] {Tikz/nlpdata_vw.dat};
\addplot+[only marks,mark size=3pt ] table[x=ns, y=wrmse] {Tikz/nlpdata_vw.dat};
\addplot+[only marks,mark size=3pt] table[x=ns, y=vmcsrmse] {Tikz/nlpdata_vw.dat};
\addplot+[only marks,mark size=3pt ] table[x=ns, y=wmcsrmse] {Tikz/nlpdata_vw.dat};
\addplot+[only marks,mark size=3pt ] coordinates {
	(400, 72)
};
\addplot+[only marks,mark size=3pt ] coordinates {
	(400,  71.2)
};
\addplot+[only marks,mark size=3pt ] coordinates {
	(1600,  70.9)
};
\legend{V-MLMLP, W-MLMLP,V-MLMLPCSS, W-MLMLPCSS, B-MLMLP, B-MLMLPCSS, Gibbs}
\end{axis}
\end{tikzpicture}}
		\caption{nlpdata}
		\label{fig:vwcycle_nlpdata}
	\end{subfigure}
	\tikzsetnextfilename{chembl_vwcycle}
	\begin{subfigure}{0.45\textwidth}
		\centering
		\resizebox{\linewidth}{!}{\begin{tikzpicture}
\begin{axis}[
height=8cm,
width=12cm,
xmode = log,
cycle list name=clusters,
ymode = log,
grid=both,
grid style={line width=.1pt, draw=gray!20},
major grid style={line width=.2pt,draw=gray!60},
minor tick num=5,
 log ticks with fixed point,
 ytick={15,17,19,21,23,25},
xtick={3,10,30,100,400,1600},
xticklabels={3,10,30,100,Consec.,All fine},
xlabel=Number of samples per change of level,ylabel=RMSE,
legend pos=north east,
legend cell align={left},
legend style={font=\small}
]
\addplot+[only marks,mark size=3pt] table[x=ns, y=vrmse] {Tikz/chembl_vw.dat};
\addplot+[only marks,mark size=3pt ] table[x=ns, y=wrmse] {Tikz/chembl_vw.dat};
\addplot+[only marks,mark size=3pt] table[x=ns, y=vmcsrmse] {Tikz/chembl_vw.dat};
\addplot+[only marks,mark size=3pt ] table[x=ns, y=wmcsrmse] {Tikz/chembl_vw.dat};
\addplot+[only marks,mark size=3pt ] coordinates {
	(400,  15.7)
};
\addplot+[only marks,mark size=3pt ] coordinates {
	(400,  15.6)
};
\addplot+[only marks,mark size=3pt ] coordinates {
	(1600,  15.4)
};
\legend{V-MLMLP, W-MLMLP,V-MLMLPCSS, W-MLMLPCSS, B-MLMLP, B-MLMLPCSS, Gibbs}
\end{axis}
\end{tikzpicture}}
		\caption{ChEMBL}
		\label{fig:vwcycle_chembl}
	\end{subfigure}
	\caption{RMSE given in function of different samples configurations as given in Table~\ref{tab:config_samples} for DrivFace, SNP1, nlpdata and ChEMBL using MLMLP-G and MLMLPCSS-G. The reference values for taking $H_l$ consecutively (consec.) and Gibbs sampling from Table~\ref{tab:overview} are additionally given. }
	\label{fig:vwcycle}
\end{figure*}

Figure \ref{fig:vwcycle} now shows the RMSE for the different configurations in Table~\ref{tab:config_samples}. As can be seen from Figure \ref{fig:vwcycle}, more than $10$ samples should be taken before changing to another level. It takes a few, generally less than $3$, samples for the Markov Chain to adopt to the new level. This can easily be incorporated by using a burn-in after a level change or by means of thinning. If enough samples are taken before each level change, one can achieve the same accuracy as if the samples were taken consecutively and equally distributed over the levels with the remark that when using these configurations less samples are taken on the fine level. 

As seen from Table~\ref{tab:config_samples} and Figure~\ref{fig:vwcycle}, it is possible to achieve high accuracy without taking much samples at the finest and thus expensive levels. For multilevel Markov chain Monte Carlo involving PDEs, one can determine the optimal number of samples on each level \cite{Dodwell2015Hierarchical}. The number of samples on level $l$ is given by \begin{equation}
H_l=\frac{2}{\epsilon^2}\left(\sum_{l=0}^L\sqrt{s_l^2C_l}\right)\sqrt{\frac{s_l^2}{C_l}} \label{eq:Hl}
\end{equation} 
with $C_l$ the cost to take one sample at level $l$, $s_l^2$ the sample variance at level $l$ and $\epsilon$ the required tolerance on the RMSE of the quantity of interest. 

In contrast, we are interested in the predictions for new and unknown data. We actually do not know our quantity of interest. One could optimize the number of samples based on the training set, but these are noisy observations. A closer look at equation \eqref{eq:Hl} shows that the factor $\frac{2}{\epsilon^2}\left(\sum_{l=0}^L\sqrt{s_l^2C_l}\right)$ is the same for all levels and the factor $\sqrt{\frac{s_l^2}{C_l}}$ is level specific. The latter can thus be used to define the ratio between the number of samples $H_l$ at each level $l$. Using this factor we define the number of samples at level $l$ as
\begin{equation}
	H_l=\left\lfloor\frac{\sqrt{\frac{s_l^2}{C_l}}}{\left(\sum_{l=0}^L\sqrt{\frac{s_l^2}{C_l}}\right)}H\right\rfloor \label{eq:hlvar}
\end{equation}
with $H$ the given total number samples over all levels and $s_l^2$ the sample variance of the average of the observations $\textbf{y}$. The cost $C_l$ to take one sample is defined as the number of nonzeros of $X_l$. Definition \ref{eq:hlvar} is based on the fact that the variance of the difference $\mathbb{V}[\textbf{y}_l-\textbf{y}_{l-1}]$ is smaller than the variance $\mathbb{V}[\textbf{y}_l]$ on level $l$ due to positive correlation. As an alternative we considered a ratio purely based on the cost for taking one sample for the multilevel sampler without correlated samples 
\begin{equation}
H_l=\left\lfloor\frac{\frac{1}{C_l}}{\left(\sum_{l=0}^L\frac{1}{C_l}\right)}H\right\rfloor \label{eq:hlcost}
\end{equation} with $C_l$ and $H$ as before. 

\begin{table}[ht]
	\centering
	\scalebox{0.9}{
		\begin{tabular}{ ll|c|c|c|c|c}
			& &DrivFace & nlpdata & SNP2 & news20 &  ChEMBL\\ \midrule

			&Gibbs  &1.13E3& 3.23E2 & 4.34E4& 7.22E3 &1.72E3\\
			&MLMLP & \textbf{2.40E2}&1.98E2 &8.91E3 &4.22E3&1.35E3\\
			&MLMLPCSS &3.88E2&3.35E2&1.41E4&6.75E3&2.44E3\\			
			& Var-MLMLP &2.43E2&\textbf{1.97E2}&8.80E3 &2.87E3 &1.29E3 \\
			Execution  & Cost-MLMLP&2.49E2&2.09E2&\textbf{4.51E3} &8.89E3 &1.30E3\\
			 time(s)& Var-MLMLPCSS&3.05E2&2.29E2& 9.86E3&\textbf{1.87E3} &\textbf{1.00E3}\\
			& Cost-MLMLPCSS &3.83E2 &3.41E2&1.25E4 &7.00E3 &2.30E3\\ \midrule
			
			&Gibbs  & 9.41&7.09E1 &1.57E-1&2.87E1 & 1.54E1\\
			&MLMLP  &9.36&7.20E1&\textbf{1.52E-1 }&3.10E1&1.57E1	\\
			&MLMLPCSS&9.40&7.12E1&1.54E-1 &3.17E1&\textbf{1.56E1}	\\			
			& Var-MLMLP &9.42&7.78E1&1.53E-1 &3.32E1 &\textbf{1.56E1} \\
			RMSE & Cost-MLMLP&\textbf{9.32}&7.19E1&\textbf{1.52E-1} &\textbf{3.06E1} &\textbf{1.56E1}\\
			& Var-MLMLPCSS&9.49&7.25E1&1.55E-1 &3.30E1 &2.09E1\\
			& Cost-MLMLPCSS& 9.40& \textbf{7.08E1} &1.54E-1 &3.32E1 &\textbf{1.56E1}\\ 					
			\bottomrule
	\end{tabular}}
	\caption{The average execution time (s)and RMSE using different data sets for multilevel sampling with and without correlated samples with the number of samples $H_l$ based on the variance (Var) of the levels in equation \eqref{eq:hlvar} (Var) or the number of nonzeros of $X_l$ (Cost) in equation \eqref{eq:hlcost}. The values for Gibbs, MLMLP-G and MLMLPCSS with the samples equally distributed over levels from Table~\ref{tab:overview} are provided as reference. The best values for the multilevel sampling schemes are presented in bold. }
		\label{tab:hl_estimation}%
\end{table}

Table~\ref{tab:hl_estimation} details the average execution time and the RMSE for the multilevel sampler with and without correlated samples with the number of samples $H_l$ for each level $l$ determined by \eqref{eq:hlvar} or \eqref{eq:hlcost}. Using the variance of the average predictions $\textbf{y}$ on level $l$ combined with the correlated samples performs worse than using only the cost of taking one sample on level $l$. The sample variance was calculated using $50$ samples on each level and then the number of total samples on each level was calculated using \eqref{eq:hlvar}. Using the cost based distribution of number of samples $H$ in equation \eqref{eq:hlcost} results in more samples at the coarser levels and generally without loss in accuracy with respect to uniformly distributing the number of samples $H$ over the levels in MLMLP. This results in slightly faster execution time.   
\subsection{Correlated samples within the Gibbs sampler}
As shown in previous results, using correlated samples can improve accuracy but does increase the execution time since each sample requires two linear solves on the adjacent levels. As mentioned in Section \ref{sect:correlation}, there is a second way to define correlation between samples. The second approach simply projects the finer solution to the coarser level. This could lead to problems when sampling from different distributions for the same level depending on the quality of the clustering. If all features within one cluster were the same, this would not be a problem. In reality this is almost never true. Table~\ref{tab:mcsmcpcomp} shows the execution time, pearson correlation and RMSE for correlated samples using two linear solves (CSS) and projecting the fine level solution to the coarser level (CSP). For almost all data sets, the accuracy of CSP, although faster, is significantly less than CSS.    

\begin{table}[ht]
	\centering
	\scalebox{0.8}{
	\begin{tabular}{ l|ccc|ccc|ccc}
		&\multicolumn{3}{c|}{Execution  time(s)}&\multicolumn{3}{c|}{Pearson Correlation}&\multicolumn{3}{c}{RMSE}\\
		Data&Gibbs& CSS& CSP&Gibbs& CSS& CSP &Gibbs& CSS& CSP    \\ \midrule			
		DrivFace&1.13E3&3.88E2&2.65E2&0.974&0.974&0.974&9.41&9.40&9.50\\			
		nlpdata&3.23E2&3.35E2&2.07E2&0.955&0.955&0.955&7.09E1&7.12E1&7.11E1\\					
		SNP2&4.34E4&9.43E3&6.45E3&0.959&0.960&0.958&1.57E-1&1.54E-1&1.60E-1\\
		news20&7.22E3&6.75E3&4.25E3&0.912&0.892&0.859&2.87E1&3.17E1&3.68E1\\				
		E2006&7.76E3&2.85E3&1.77E3&0.962&0.961&0.957&7.88E-2&8.01E-2&8.33E-2\\
		ChEMBL&1.72E3&2.44E3&1.39E3&0.772&0.768&0.766&1.54E1&1.56E1&1.57E1\\		
		\bottomrule
	\end{tabular}}
\caption{The average execution time (s), pearson correlation and RMSE using different data sets for Gibbs sampling and multilevel sampling using correlated samples by linear solves (CSS) and Projections (CSP). Using CSP performs worse than CSS for almost all data sets. }
\label{tab:mcsmcpcomp}%
\end{table}

\begin{figure*}[ht]
	\centering
	\tikzsetnextfilename{nlpdata_lvl_var}
	\begin{subfigure}{0.45\textwidth}
		\centering
		\resizebox{\linewidth}{!}{\begin{tikzpicture}
\begin{axis}[
height=8cm,
width=12cm,
cycle list name=clusters,
ymode = log,
grid=both,
grid style={line width=.1pt, draw=gray!20},
major grid style={line width=.2pt,draw=gray!60},
minor tick num=5,
xtick={0,1,2},
xlabel=$l$,ylabel=$\mathbb{V}$,
legend pos=south west,
legend cell align={left},
legend style={font=\small}
]

\addplot+[only marks,mark size=3pt] table[x=lvl, y=var_lvls] {Tikz/nlpdata_lvl_var.dat};
\addplot+[only marks,mark size=3pt] table[x=lvl, y=dvar_lvls] {Tikz/nlpdata_lvl_var.dat};
\addplot+[only marks,mark size=3pt] table[x=lvl, y=var_lvlp] {Tikz/nlpdata_lvl_var.dat};
\addplot+[only marks,mark size=3pt] table[x=lvl, y=dvar_lvlp] {Tikz/nlpdata_lvl_var.dat};
\legend{CSS  $y_l$,CSS $y_l-y_{l-1}$,CSP  $y_l$,CSP $y_l-y_{l-1}$}
\end{axis}
\end{tikzpicture}}
		\caption{nlpdata}
		\label{fig:lvlvar_nlpdata}
	\end{subfigure}
	\tikzsetnextfilename{snp2_lvl_var}
	\begin{subfigure}{0.45\textwidth}
		\centering
		\resizebox{\linewidth}{!}{\begin{tikzpicture}
\begin{axis}[
height=8cm,
width=12cm,
cycle list name=clusters,
ymode = log,
grid=both,
grid style={line width=.1pt, draw=gray!20},
major grid style={line width=.2pt,draw=gray!60},
minor tick num=5,
xtick={0,1,2,3},
xlabel=$l$,ylabel=$\mathbb{V}$,
legend pos=south west,
legend cell align={left},
legend style={font=\small}
]

\addplot+[only marks,mark size=3pt] table[x=lvl, y=var_lvls] {Tikz/snp2_lvl_var.dat};
\addplot+[only marks,mark size=3pt] table[x=lvl, y=dvar_lvls] {Tikz/snp2_lvl_var.dat};
\addplot+[only marks,mark size=3pt] table[x=lvl, y=var_lvlp] {Tikz/snp2_lvl_var.dat};
\addplot+[only marks,mark size=3pt] table[x=lvl, y=dvar_lvlp] {Tikz/snp2_lvl_var.dat};
\legend{CSS  $y_l$,CSS $y_l-y_{l-1}$,CSP  $y_l$,CSP $y_l-y_{l-1}$}
\end{axis}
\end{tikzpicture}}
		\caption{SNP2}
		\label{fig:lvlvar_snp2}
	\end{subfigure}
	\tikzsetnextfilename{E2006_lvl_var}
	\begin{subfigure}{0.45\textwidth}
		\centering
		\resizebox{\linewidth}{!}{\begin{tikzpicture}
\begin{axis}[
height=8cm,
width=12cm,
cycle list name=clusters,
ymode = log,
grid=both,
grid style={line width=.1pt, draw=gray!20},
major grid style={line width=.2pt,draw=gray!60},
minor tick num=5,
xtick={0,1,2,3},
xlabel=$l$,ylabel=$\mathbb{V} $,
legend pos=south west,
legend cell align={left},
legend style={font=\small}
]
\addplot+[only marks,mark size=3pt] table[x=lvl, y=var_lvls] {Tikz/E2006_lvl_var.dat};
\addplot+[only marks,mark size=3pt] table[x=lvl, y=dvar_lvls] {Tikz/E2006_lvl_var.dat};
\addplot+[only marks,mark size=3pt] table[x=lvl, y=var_lvlp] {Tikz/E2006_lvl_var.dat};
\addplot+[only marks,mark size=3pt] table[x=lvl, y=dvar_lvlp] {Tikz/E2006_lvl_var.dat};
\legend{CSS  $y_l$,CSS $y_l-y_{l-1}$,CSP  $y_l$,CSP $y_l-y_{l-1}$}
\end{axis}
\end{tikzpicture}}
		\caption{E2006}
		\label{fig:lvlvar_E2006}
	\end{subfigure}
	\tikzsetnextfilename{chembl_lvl_var}
		\begin{subfigure}{0.45\textwidth}
		\centering
		\resizebox{\linewidth}{!}{\begin{tikzpicture}
\begin{axis}[
height=8cm,
width=12cm,
cycle list name=clusters,
ymode = log,
grid=both,
grid style={line width=.1pt, draw=gray!20},
major grid style={line width=.2pt,draw=gray!60},
minor tick num=5,
xtick={0,1,2,3,4,5},
xlabel=$l$,ylabel=$\mathbb{V}$,
legend pos=south west,
legend cell align={left},
legend style={font=\small}
]
\addplot+[only marks,mark size=3pt] table[x=lvl, y=var_lvls] {Tikz/chembl_lvl_var.dat};
\addplot+[only marks,mark size=3pt] table[x=lvl, y=dvar_lvls] {Tikz/chembl_lvl_var.dat};
\addplot+[only marks,mark size=3pt] table[x=lvl, y=var_lvlp] {Tikz/chembl_lvl_var.dat};
\addplot+[only marks,mark size=3pt] table[x=lvl, y=dvar_lvlp] {Tikz/chembl_lvl_var.dat};
\legend{CSS  $y_l$,CSS $y_l-y_{l-1}$,CSP  $y_l$,CSP $y_l-y_{l-1}$}
\end{axis}
\end{tikzpicture}}
		\caption{ChEMBL}
		\label{fig:lvlvar_chembl}
	\end{subfigure}
	\caption{The variance $\mathbb{V}[\textbf{y}_l]$ of level $l$ for $l=0, \dots, L$ and the variance $\mathbb{V}[\textbf{y}_l-\textbf{y}_{l-1}]$ of the difference for $l=1, \dots, L$ on the training data of nlpdata, SNP2, E2006 and ChEMBL. The results are given using correlated samples with linear solves (CSS) and projections (CSP).  }
\label{fig:lvlvar}
\end{figure*}
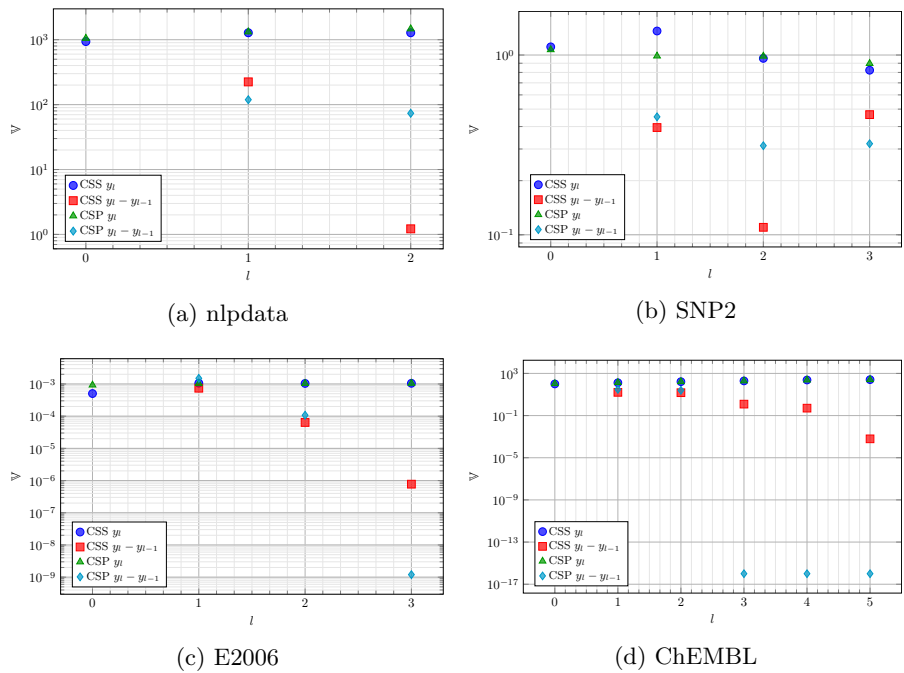

Figure~\ref{fig:lvlvar} shows the variances for each level $l$ and variance of the difference of the correlated samples for observation $\textbf{y}(0)$ for CSS and CSP. The figure shows that, as expected, the variance of the difference on level $l$  and $l-1$ is smaller than the variance of level $l$. For the ChEMBL data, the finer levels for CSP do not contain additional information with respect to the coarser levels to aid in the prediction of $\textbf{y}(0)$ resulting in zero variance.    
 
\section{Conclusion}\label{sect:conc}
We have presented a multilevel Gibbs sampler for Bayesian regression for linear mixed models. Using clustering algorithms, a hierarchy of data matrices is created. This hierarchy allows to sample on different levels reducing execution time without significant loss in predictive performance for almost all data sets. Since the coarser levels contain most of the variance of the data, more samples can be taken at the coarser levels and the chain converges faster than taking all the samples at the finest level. Distributing the number of samples over the levels based on the cost to take one level sample was shown to provide fast and accurate results. 

Furthermore, we investigated the use of correlated samples to reduce the variance of the Markov Chain. The use of correlated samples increased the execution time but did not consistently improve the accuracy of the predictions. The distribution of the number of samples across the levels based on the variance performed less optimal than distribution based on sampling cost alone. The multilevel sampler with correlated samples was often still faster than plain sampling on the fine level and by storing both the solutions of level $l$ and the difference between the levels $l$ and $l-1$, it is possible to get the predictions of both multilevel samplers as part of an ensemble.  
\section{Acknowledgement}		
This work was supported Research Foundation - Flanders (FWO, No. G079016N). We would like to thank David Vanavermaete for generating the SNP data sets. 		
\bibliographystyle{siamplain}
\bibliography{../references} 
	
\end{document}